\documentclass{article}

\usepackage{multicol}
\usepackage[utf8]{inputenc}
\usepackage[T1]{fontenc}
\usepackage[a4paper,
            bindingoffset=5mm,
            left=15mm,
            right=15mm,
            top=15mm,
            bottom=15mm,
            footskip=5mm]{geometry}
\usepackage{amssymb}
\usepackage{mathtools}
\usepackage{graphicx}
\usepackage{verbatim}
\usepackage{xcolor}
\usepackage{tabularx}
\usepackage{subcaption}
\usepackage[hidelinks]{hyperref}
\usepackage{float}
\usepackage{physics}
\usepackage{siunitx}
\usepackage{cite}

\newcommand{\keywords}[1]{ \small \textbf{\textit{Keywords:}} #1}
\newcommand{\boldlongemdash}{\textbf{\fontfamily{lmss}\selectfont---}}
\title{Plasma Position Constrained Free-Boundary MHD Equilibrium in Tokamaks using pyIPREQ}
\author{Udaya Maurya$^{1,2,a}$, Amit K. Singh$^{1,2,3}$, Suman Aich$^{1,2}$, Jagabandhu Kumar$^{1,2}$, \\ Rohit Kumar$^1$, Daniel Raju$^{1,2}$ \\ \small $^1$Institute for Plasma Research, Bhat, Gandhinagar, India, 382428 \\ \small $^2$Homi Bhabha National Institute, Anushaktinagar, Mumbai, India, 400094 \\ \small $^3$ ITER-India, Institute for Plasma Research, Bhat, Gandhinagar 382428, Gujarat, India \\ \small $^a$ Author to whom correspondence should be addressed: udaya.maurya@ipr.res.in}
\hypersetup{pdfstartview={XYZ null null 1.00},
                      pdftitle={Plasma Position Constrained Free-Boundary MHD Equilibrium in Tokamaks using pyIPREQ},
                      pdfauthor={Udaya Maurya}}

\begin{document}
\maketitle
\begin{center}
\begin{tabular}{@{} p{0.85\columnwidth} @{}}
\begin{abstract}
A free-boundary, axisymmetric magnetohydrodynamic (MHD) equilibrium code, pyIPREQ, has been developed for Tokamak plasmas using finite difference and Green's function approach. The code builds upon the foundational frameworks of the PEST and IPREQ codes, introducing several enhancements and new capabilities. Notably, pyIPREQ supports the specification of limiter boundaries and enables the computation of key physical quantities. The code has also been extended to compute equilibria constrained by a prescribed magnetic axis position, which is particularly useful when such information can be inferred from the diagnostics data. In addition, pyIPREQ includes functionality to address vertical instabilities, a requirement for accurately modeling elongated plasma configurations. Benchmarking has been carried out against published results and the original IPREQ code. Applications are demonstrated for ADITYA-U Tokamak experiments, where magnetic axis measurements are available, and predictions are also made for SST-1 and ADITYA-U Tokamaks under various operational scenarios.
\end{abstract}
\end{tabular}
\end{center}
\hspace{0.12\columnwidth}\keywords{MHD equilibrium, Grad-Shafranov equation, free-boundary,\\ \phantom{\keywords}\hspace{0.1225\columnwidth}ITER, ADITYA-U, SST-1}

\begin{multicols}{2}

\section{Introduction}
\label{sec:intro}

In a Tokamak, magnetohydrodynamic (MHD) equilibrium is established through the balance between the plasma thermal pressure and the total magnetic pressure. Studying MHD equilibrium is essential in Tokamak plasma discharge experiments because it helps in determining plasma shape and stability, interpreting and diagnosing experimental observations, estimating key physical parameters, and gaining insights into plasma confinement and transport. This equilibrium condition is governed by the Grad-Shafranov equation, which is an elliptic partial differential equation and can be derived by equating thermal pressure and magnetic pressure, with axisymmetry. Consider a cylindrical coordinate system ($R$, $\phi$, $Z$) where $R$ represents the radial distance from the Tokamak's central axis, $\phi$ is the toroidal angle, and $Z$ denotes the vertical coordinate. The Grad-Shafranov equation in terms of poloidal magnetic flux $\psi$, poloidal current function $F$ (related with toroidal magnetic field $B_{\phi}$ as $F=RB_{\phi}$) and thermal pressure $p$, can be written in SI units as:
\begin{equation}
\frac{\partial^2\psi}{\partial R^2}+\frac{\partial^2\psi}{\partial Z^2}-\frac{1}{R}\frac{\partial\psi}{\partial R}=-\mu_0R^2\frac{\partial p}{\partial\psi}-F\frac{\partial F}{\partial\psi}
\label{eq:gs}
\end{equation}
The free-boundary problem for a Tokamak plasma discharge operation requires solving equation \ref{eq:gs} to find $\psi$, while incorporating the effects of both plasma and external poloidal field coils. The axisymmetry assumption implies symmetry in $\phi$ direction and solving the problem is reduced in $R$-$Z$ plane, called the poloidal plane. The left-side of equation \ref{eq:gs} is often written as $\Delta^*\psi$ (where $\Delta^*$ is known as the Grad-Shafranov operator) and right-side is related to the toroidal current density $J_\phi$ as
\begin{equation}
J_{\phi}=Rp'+\frac{1}{\mu_0R}FF'
\label{eq:Jphi}
\end{equation}
where $p'=\partial p/\partial\psi$ and $FF'=F\partial F/\partial\psi$. Consequently, equation \ref{eq:gs} can also be written as $\Delta^*\psi=-\mu_0RJ_{\phi}$. \\

The IPREQ code \cite{2020Sharma}, which was inspired by PEST code \cite{1979Johnson}, solves the free-boundary problem with given external coil currents and basic Tokamak machine parameters. In addition, the variables $p'$ and $FF'$ in right-side of equation \ref{eq:gs} are assumed to have some profile in $\psi$, which is often necessary due to lack of proper Tokamak diagnostics data. IPREQ and PEST codes also solve the fixed-boundary problem, where coil currents are estimated based on a given plasma boundary, however this feature is out of scope of this article. Some free-boundary codes like EFIT \cite{1985Lao}, FBT \cite{1988Hofmann}, CLISTE  \cite{1999McCarthy}, LIUQE \cite{2015Moret}, NICE \cite{2020Faugeras}, etc. solve the reconstruction problem by optimizing $p'$ and $FF'$ profiles to best match the experimental diagnostic data. The pyIPREQ code, developed in Python-3 programming language and presented in this article, solves the free-boundary MHD equilibrium problem similar to IPREQ. It has also been extended to determine $\psi$ while enforcing a known magnetic axis position, which can be inferred from various Tokamak diagnostics data. The underlying algorithms are discussed in detail in sections \ref{sec:algo0}, \ref{sec:algo1} and \ref{sec:algo1Z}. The validation of pyIPREQ with a published ITER result and the original IPREQ code is presented in section \ref{sec:valid0}. For the magnetic axis position constrained part of the pyIPREQ, results for some Tokamak plasma discharge experiments are discussed in sections \ref{sec:testRax}, \ref{sec:results1} and \ref{sec:algo1Z}. Finally, to demonstrate the applications of the code, some predictions are made for proposed SST-1 upgrades and a previous ADITYA-U experiment in sections \ref{sec:predsst1u} and \ref{sec:predadityau}.

\section{Free-boundary MHD Equilibrium}
The pyIPREQ code generally follows the mathematical and numerical frameworks outlined in \cite{1979Johnson}, with certain enhancements and extensions. Section \ref{sec:algo0} provides a brief overview of the core numerical methods and procedures implemented in the pyIPREQ code, followed with section \ref{sec:valid0} discussing the validation.
\subsection{Algorithm}
\label{sec:algo0}
The objective of solving the free-boundary problem is to obtain $\psi$ in the poloidal plane. This plane is numerically described by a uniform rectangular $R$-$Z$ grid. This grid should entirely cover the plasma domain and can optionally include external poloidal field coils. First, $\psi_{bc}$ is computed, which is $\psi$ due to external coils on the grid boundary points $(R_k,Z_k)$. Since right-side of equation \ref{eq:gs} is 0 in vacuum, the equation becomes $\Delta^*\psi=0$, which can be easily solved using the Green's function method:
\begin{equation}
\begin{split}
& \psi_{bc}(R_k,Z_k)=\frac{\mu_0}{\pi}\sum_{j=1}^{n_{coils}}\sum_{i=1}^{n_j}I_iG(R_k,Z_k,R_i,Z_i) \\
& G(R_k,Z_k,R_i,Z_i)=\frac{\sqrt{R_kR_i}}{\kappa} \\
& \phantom{G(R_k,Z_k,R_i,Z_i)}\left[\left(1-\frac{\kappa^2}{2}\right)K\left(\kappa^2\right)-E\left(\kappa^2\right)\right] \\
& \kappa^2=\frac{4R_kR_i}{\left(R_k+R_i\right)^2+\left(Z_k-Z_i\right)^2}
\end{split}
\label{eq:psibc}
\end{equation}
where $G$ is the Green's function of the Grad-Shafranov operator, $K\left(x\right)$ and $E\left(x\right)$ are complete elliptic integrals of first and second kind defined as $K\left(x\right)=\int_0^{\pi/2}d\theta/\sqrt{1-x\sin^2\theta}$ and $E\left(x\right)=\int_0^{\pi/2}d\theta\sqrt{1-x\sin^2\theta}$. Usually, the external coils are further divided in smaller \emph{filaments} when computing $\psi_{bc}$ and current in each coil is equally distributed among these filaments, the index $i$ in equation \ref{eq:psibc} represents each filament. The index $j$ corresponds to external coils and $n_j$ is the number of filaments of the $j^{th}$ coil. This filamentation of coils should be fine enough so that error due to this discretization does not significantly impact the equilibrium solution. Following similar approach, $\psi$ at the boundary due to plasma, $\psi_{bp}$, is computed as:
\begin{equation}
\psi_{bp}(R_k,Z_k)=\frac{\mu_0}{\pi}\Delta R\,\Delta Z\sum_{i=1}^{n_R}\sum_{j=1}^{n_Z}J_{\phi,ij}G(R_k,Z_k,R_i,Z_j)
\label{eq:psibp}
\end{equation}
where sums are over entire $R$-$Z$ grid, $\Delta R, \Delta Z$ refer to grid spacing and $n_R,n_Z$ refer to number of grid points. $J_\phi$ is assumed to have a starting profile of the form $\left(1-r^2/a^2\right)$, where $r$ is the radius from the major radius $R_0$ and $a$ is the minor radius. This initial $J_\phi$ is constrained within limiter configuration so that $J_\phi=0$ outside radius $a$. The magnitude of $J_\phi$ is constrained with given plasma current $I_p$ such that $\iint J_\phi\,dR\,dZ=I_p$. \\

Next, $\psi$ is computed on the rest of the $R$-$Z$ grid using finite-difference method with $\psi_b=\psi_{bc}+\psi_{bp}$ as the Dirichlet boundary conditions. The derivatives in left-side of equation \ref{eq:gs} are approximated using central difference formula and the resultant set of linear equations can be directly solved. This newly obtained $\psi$ is then used to compute a new $J_\phi$, which is assumed to have a certain profile compatible with \ref{eq:Jphi}, for example:
\begin{equation}
J_\phi\propto\left(\beta\frac{R}{R_0}+\left(1-\beta\right)\frac{R_0}{R}\right)\left(1-\left(1-\bar\psi\right)^\alpha\right)^\gamma
\label{eq:Jprofile}
\end{equation}
where $\bar\psi$ is normalized $\psi$ such that $\bar\psi$ has maxima $1$ (that point is called the magnetic axis) and goes to $0$ at the plasma boundary (where $\psi$ first touches the limiter configuration). $\alpha$, $\beta$ and $\gamma$ are input profile parameters, which can be either guessed or estimated using experimental diagnostic data. The magnitude of $J_\phi$ is kept constrained with plasma current $I_p$ and $J_\phi$ is truncated after $\bar\psi=0$. A different profile of $J_\phi$ other than the equation \ref{eq:Jprofile} can also be assumed as long as it has the same form of equation \ref{eq:Jphi}. $\psi$ is recomputed using this new $J_\phi$, and they are iteratively computed until a convergence is obtained resulting in a self-consistent solution of the Grad-Shafranov equation. Such iterations are often used in root-finding algorithms and are called Picard iterations.

A typical problem that arises in solving the free-boundary problem for elongated plasmas is numerical vertical instability. This instability arises because during the iterations $J_\phi$ and $\psi$ could be such that there is a net vertical force $\left(\vec{J}\times\vec{B}\right)_Z$, even though such a force may not be present in the final solution. During the iterations such a force can drift the solution towards the edge. This instability is identifiable by the parameter $n_{\text{decay}}\equiv \left(R/B_Z\right)\partial B_Z/\partial R$. It can occur in a vertically elongated plasma if $n_{\text{decay}}<0$. A similar horizontal instability can occur in a horizontally elongated plasma if $n_{\text{decay}}>1.5$ \cite{2015Jeon}. To resolve this problem, the solution by \cite{1979Johnson} is adapted here. Two artificial feedback coils are introduced with horizontal position at the center and vertical position just outside the upper and lower boundaries of the computational domain. Then the following feedback current is provided in these coils at each step of the iteration, with a desired vertical magnetic axis position $Z_{\text{target}}$.
\begin{equation}
I_{\text{coil}}=-\text{sgn}\left({Z_{\text{coil}}}\right)C_1\left(Z_{ax}-Z_{\text{target}}\right)I_p
\label{eq:Ifbvert}
\end{equation}
where $C_1$ is an arbitrary parameter that can be found with trial-and-error (same as mentioned in \cite{1979Johnson}), $Z_{\text{coil}}$ represents the feedback coil's vertical position and $Z_{ax}$ represents the vertical position of the magnetic axis. In a successful execution of the program, $I_{\text{coil}}$ is expected to become negligible, thus nullifying the effects of these artificial coils.

Steps of the algorithm can be summarized as following:
\begin{enumerate}
\item Define uniform rectangular $R$-$Z$ grid.
\item Compute $\psi_{bc}$ using equations \ref{eq:psibc}.
\item Start with a reasonable guess profile of $J_\phi$, constrained to $I_p$ and limiters.
\item Compute $\psi_{bp}$ using equation \ref{eq:psibp}.
\item Compute $\psi$ using finite-difference method with $\psi_b$ as boundary conditions.
\item Compute $J_\phi$ using a reasonable profile in $\psi$, like equation \ref{eq:Jprofile}.
\item Repeat steps 4-6 until convergence is achieved.
\end{enumerate}
If equation \ref{eq:Ifbvert} is used for vertical stability, it needs to be computed before each computation of $\psi$, i.e., between Steps 4 and 5.

With the conventions mentioned so far, the COCOS number \cite{2013Sauter} of pyIPREQ is 7, with the interpretation that positive values of $I_p$ and $B_\phi$ are in counter-clockwise direction from top-view of the Tokamak.
\subsection{Validation}
\label{sec:valid0}
pyIPREQ has been compared with a published result for ITER \cite{2004Portone} and previously developed IPREQ code \cite{2020Sharma}. Computation of parameters like safety factor $q$, plasma beta $\beta$ and internal inductance $\ell_i$ have been compared with FREEGS code \cite{2024Ben}.
\subsubsection{Comparison with ITER case}
\begin{figure}[H]
\begin{center}
\includegraphics[width=0.505\columnwidth]{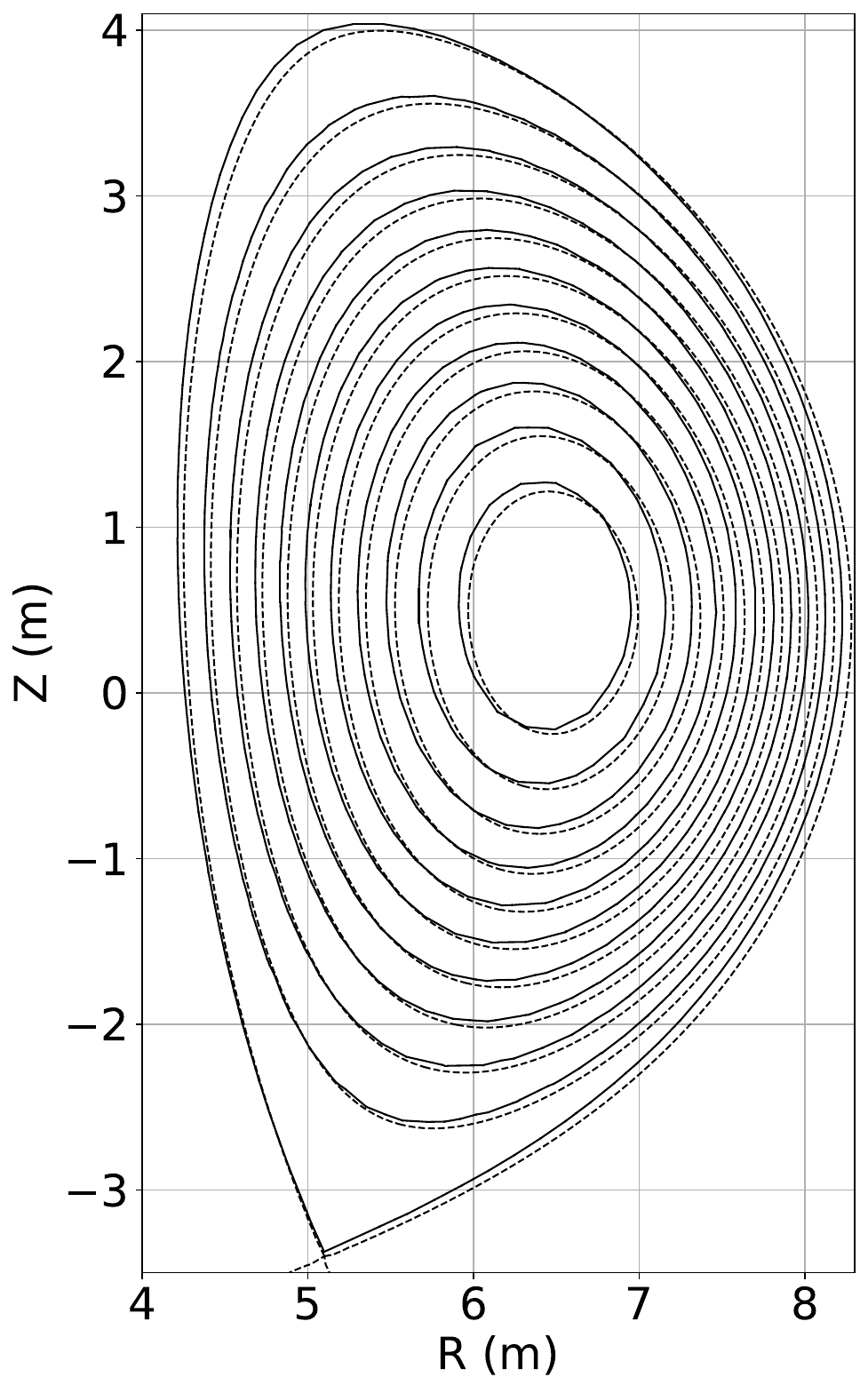}
\includegraphics[width=0.475\columnwidth]{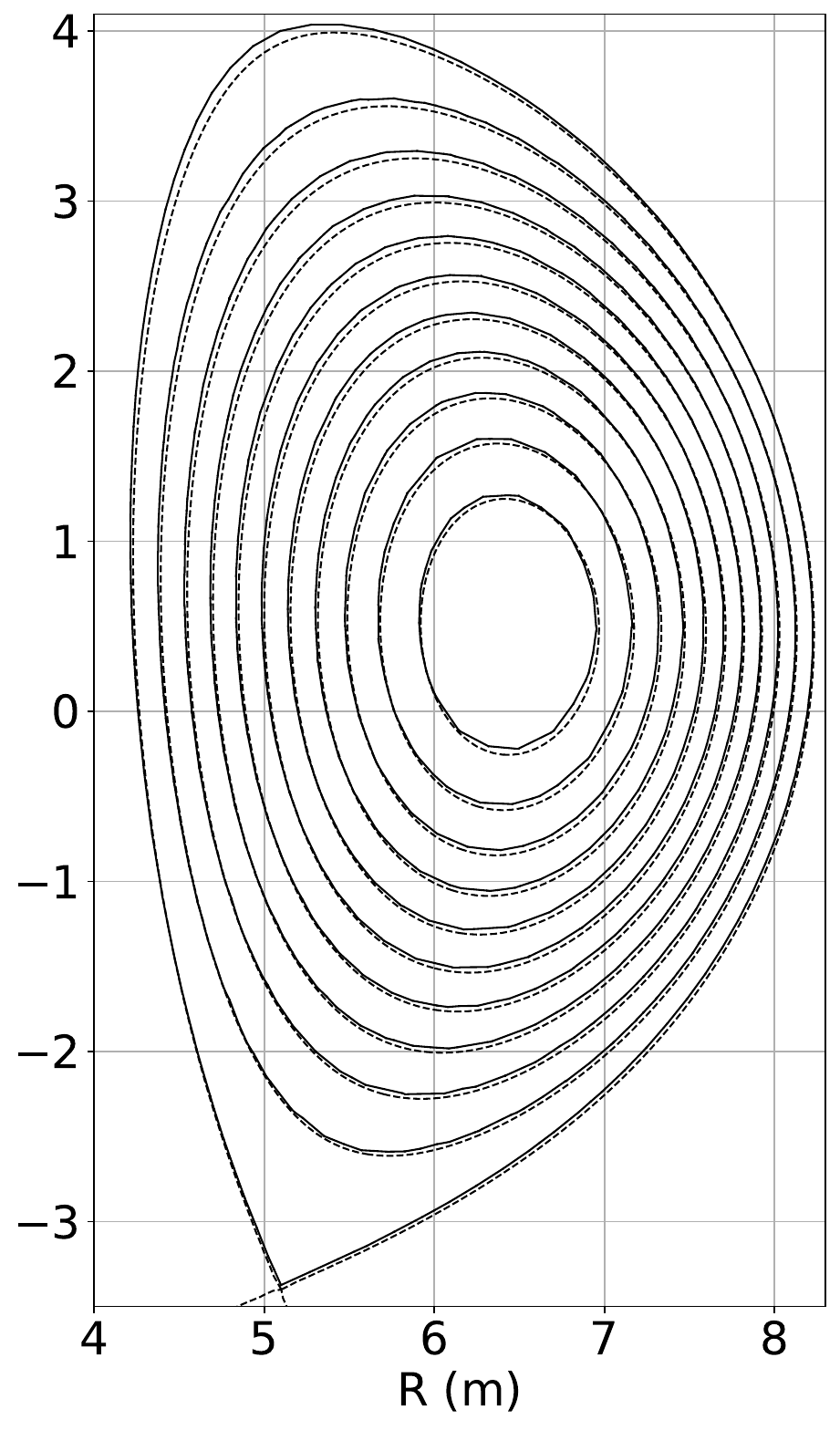}
\end{center}
\caption{Comparison of $\psi$ contours from pyIPREQ (dashed lines) with \cite{2004Portone} (solid lines), using Method-1 (left) and Method-2 (right).}
\label{fig:iterpsi}
\end{figure}

\begin{figure}[H]
\begin{center}
\includegraphics[width=0.8\columnwidth]{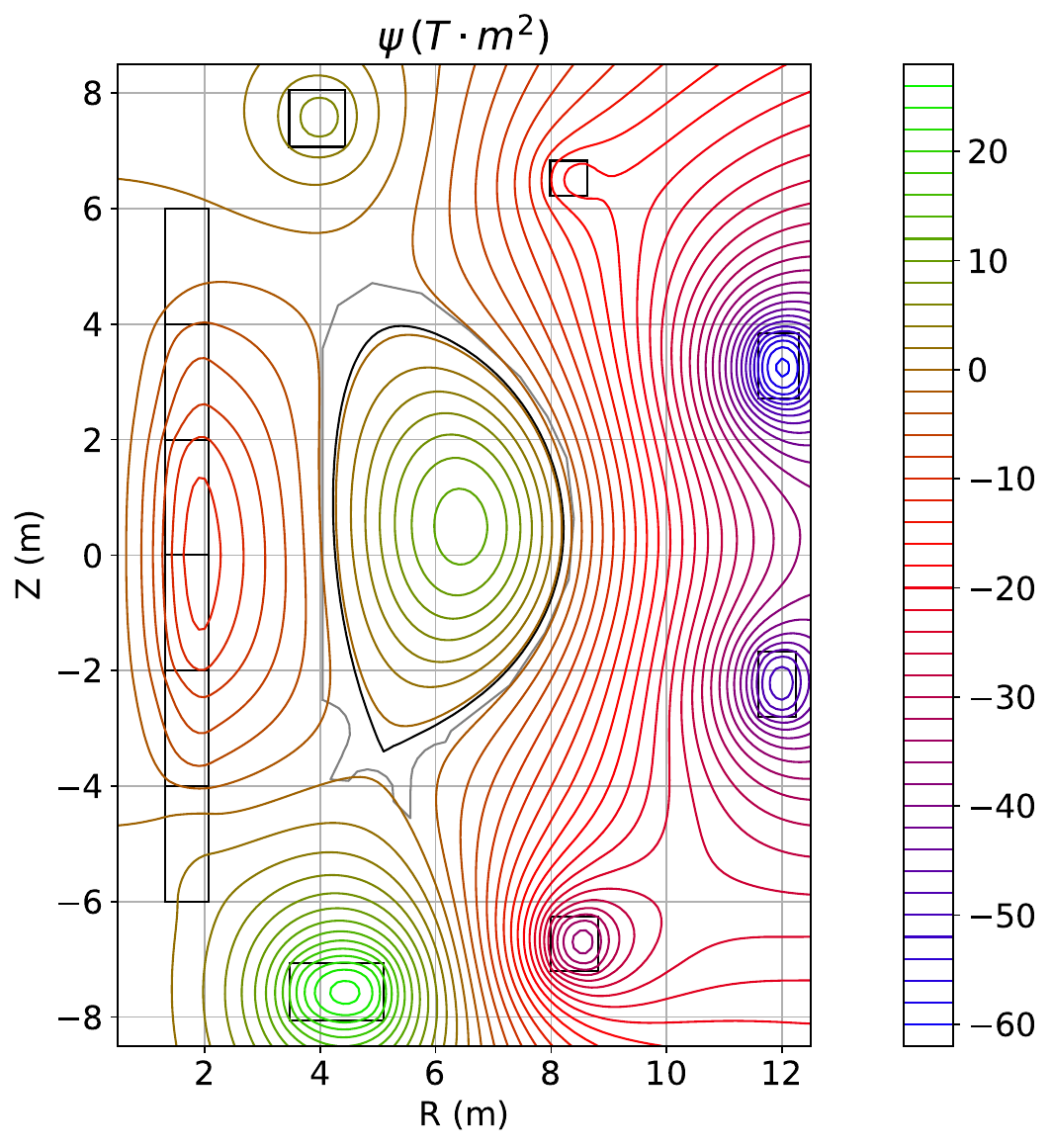}
\end{center}
\caption{$\psi$ for ITER with Method-2 on a larger grid.}
\label{fig:iterfull}
\end{figure}

In this section, a comparison of pyIPREQ with a published ITER case \cite{2004Portone} has been discussed. Following are the parameters used in this case: major radius $R_0=\SI{6.2}{\meter}$, plasma current $I_p=\SI{15.0}{MA}$, toroidal magnetic field $B_{\phi0}=\SI{5.3}{\tesla}$, current density profile $J_\phi=\lambda\left[0.63R/R_0+0.37R_0/R\right]\bar{\psi}^{0.95}$. The poloidal field coils data used is shown in table \ref{tbl:ITERcoils}.
\begin{table}[H]
\footnotesize
\begin{center}
\begin{tabular}{ l|r|r|r|r|r }
 \hline
Coil & $R\,(m)$ & $Z\,(m)$ & $\Delta R\,(m)$ & $\Delta Z\,(m)$ & $I_{\text{c0}}\,(MA)$\\
 \hline
CS3U & \num{1.69} & \num{5.00} & \num{0.75} & \num{2.00} & \num{-0.40} \\
CS2U & \num{1.69} & \num{3.00} & \num{0.75} & \num{2.00} & \num{-9.68} \\
CS1U & \num{1.69} & \num{1.00} & \num{0.75} & \num{2.00} & \num{-20.09} \\
CS1L & \num{1.69} & \num{-1.00} & \num{0.75} & \num{2.00} & \num{-20.09} \\
CS2L & \num{1.69} & \num{-3.00} & \num{0.75} & \num{2.00} & \num{-9.50} \\
CS3L & \num{1.69} & \num{-5.00} & \num{0.75} & \num{2.00} & \num{3.22} \\
PF1 & \num{3.95} & \num{7.56} & \num{0.97} & \num{0.98} & \num{4.91} \\
PF2 & \num{8.31} & \num{6.53} & \num{0.65} & \num{0.60} & \num{-2.04} \\
PF3 & \num{11.94} & \num{3.27} & \num{0.71} & \num{1.13} & \num{-6.52} \\
PF4 & \num{11.91} & \num{-2.24} & \num{0.65} & \num{1.13} & \num{-4.69} \\
PF5 & \num{8.40} & \num{-6.73} & \num{0.82} & \num{0.95} & \num{-7.54} \\
PF6 & \num{4.29} & \num{-7.56} & \num{1.63} & \num{0.98} & \num{17.2} \\
 \hline
\end{tabular}
\end{center}
\caption{Poloidal field coils used in constructing MHD equilibrium for ITER case, from $I_{c0}$ case of Table 2 of \cite{2004Portone}.}
\label{tbl:ITERcoils}
\end{table}
\normalsize
It is a highly elongated case, posing a challenge in constructing the flux surfaces and requires vertical instability resolution mentioned in section \ref{sec:algo0}. For comparison, pyIPREQ is executed with 2 different methods. In Method-1, only equation \ref{eq:Ifbvert} is used for vertical instability. In Method-2, a similar horizontal setup is also used, in addition to equation \ref{eq:Ifbvert}. In principle, there is no need to address the horizontal instability here but these additional coils produce closer results and the final currents in them are comparable with the precision of external coil currents mentioned in \cite{2004Portone}. The direct comparison of both methods is shown in figure \ref{fig:iterpsi}.
Comparison of the physical quantities internal inductance $\ell_i$, plasma poloidal beta $\beta_p$, magnetic axis position $\left(R_{ax},Z_{ax}\right)$, elongation $\kappa$, triangularity $\delta$ and factor of $J_\phi$ profile $\lambda$ are shown in table \ref{tbl:ITERvars}. The currents $I_{fbv}$ and $I_{fbh}$ are the final coil currents in the artificial feedback vertical and horizontal coils.  These coils were positioned $\SI{0.5}{\meter}$ outside of the computational domain $\left(\SI{3.5}{\meter},\SI{8.5}{\meter}\right)\times\left(\SI{-5}{\meter},\SI{5}{\meter}\right)$. While their currents could not become zero at the final step of the iteration, their values are comparable to the precision of the external coil currents mentioned in \cite{2004Portone}, i.e., \SI{0.01}{MA}.
\begin{table}[H]
\footnotesize
\begin{center}
\begin{tabular}{ l|r|r|r }
 \hline
 & From \cite{2004Portone} & Method-1 & Method-2 \\
 \hline
$\ell_i$ & \num{0.86} & \num{0.8665} & \num{0.8638} \\
$\beta_p$ & \num{0.68} & \num{0.6560} & \num{0.6510} \\
$R_{ax}$ (m) & \num{6.45} & \num{6.4949} & \num{6.4591} \\
$Z_{ax}$ (m) & \num{0.50} & \num{0.4833} & \num{0.4961} \\
$\kappa$ & \num{1.85} & \num{1.8359} & \num{1.8443} \\
$\delta$ & \num{0.48} & \num{0.4934} & \num{0.4821} \\
$\lambda/\mu_0$ (MA/m$^2$) & \num{1.66} & \num{1.6485} & \num{1.6658} \\
$I_{fbv}$ (MA) & & \num{0.1} & \num{0.066} \\
$I_{fbh}$ (MA) & & & \num{-0.044} \\
 \hline
\end{tabular}
\end{center}
\caption{Comparison of physical quantities obtained from \cite{2004Portone} and from pyIPREQ.}
\label{tbl:ITERvars}
\end{table}
The full plot of $\psi$ obtained with Method-2 on a larger grid is shown in figure \ref{fig:iterfull}.

\subsubsection{Comparison with IPREQ code}
\label{sec:IPREQcomp}
pyIPREQ has been benchmarked with IPREQ code \cite{2020Sharma} by generating two hypothetical cases of ADITYA-U \cite{2020Sharma,2024Singh,2024Tanna} and SST-1 \cite{2000Saxena,2015Pradhan,2022Sharma} Tokamaks. Instead of $J_\phi$ profile mentioned in equation \ref{eq:Jprofile}, following profiles are used:
\begin{equation}
\begin{split}
& J_\phi\propto Rf_p+\left(\frac{1}{\beta}-1\right)\frac{R_0^2}{R}f_F \\
& f_p=\frac{\exp\left(-\alpha_p\left(1-\bar\psi\right)\right)-\exp\left(-\alpha_p\right)}{\exp\left(-\alpha_p\right)-1} \\
& f_F=\frac{\exp\left(-\alpha_F\left(1-\bar\psi\right)\right)-\exp\left(-\alpha_F\right)}{\exp\left(-\alpha_F\right)-1} \\
\end{split}
\end{equation}
The basic machine parameters and poloidal field coils configuration used for these 2 cases are shown in tables \ref{tbl:IPRparams} and \ref{tbl:IPRcoils}.
\begin{table}[H]
\begin{center}
\begin{tabular}{ l|l|l }
 \hline
 & ADITYA-U & SST-1 \\
 \hline
$R_0$, $a$ & \SI{0.75}{\meter}, \SI{0.25}{\meter} & \SI{1.1}{\meter}, \SI{0.25}{\meter} \\
$I_p$ & \SI{55}{\kA} & \SI{80}{\kA} \\
$\alpha_p$, $\alpha_F$, $\beta$ & \num{0.8}, \num{1.2}, \num{0.2} & \num{0.7}, \num{1.7}, \num{0.23} \\
 \hline
\end{tabular}
\end{center}
\caption{Basic machine parameters for ADITYA-U and SST-1, used for validation of pyIPREQ.}
\label{tbl:IPRparams}
\end{table}
\begin{table}[H]
\footnotesize
\begin{center}
\begin{tabular}{ l|r|r|r|r|r }
 \hline
Coil & $R\,(m)$ & $Z\,(m)$ & $\Delta R\,(m)$ & $\Delta Z\,(m)$ & $I_{\text{coil}}\,(kA)$\\
 \hline
\multicolumn{6}{l}{ADITYA-U} \\
\hline
BV1 & \num{0.382} & $\pm$\num{1.05} & \num{0.164} & \num{0.121} & \num{-120} \\
BV2 & \num{1.64} & $\pm$\num{1.188} & \num{0.1798} & \num{0.04} & \num{-44} \\
\hline
\multicolumn{6}{l}{SST-1} \\
 \hline
 VF & \num{2.5} & $\pm$\num{1.381} & \num{0.117} & \num{0.095} & \num{-78} \\
 \hline
\end{tabular}
\end{center}
\caption{Poloidal field coils configuration for ADITYA-U and SST-1, used for validation of pyIPREQ.}
\label{tbl:IPRcoils}
\end{table}

\begin{figure}[H]
\begin{center}
\includegraphics[width=\columnwidth]{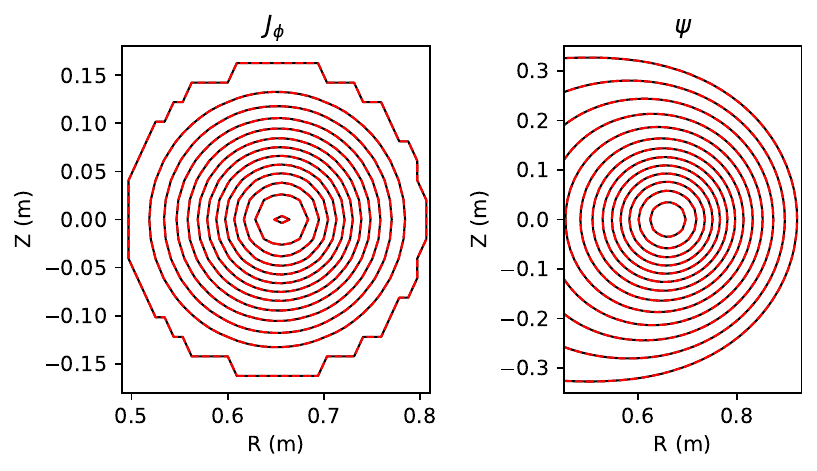}
\end{center}
\caption{Comparison of $J_\phi$ and $\psi$ from IPREQ ({\color{black}\boldlongemdash}) and pyIPREQ ({\color{red}\textbf{-\,-}}) for ADITYA-U case.}
\label{fig:validADITYA0}
\end{figure}

\begin{figure}[H]
\begin{center}
\includegraphics[width=\columnwidth]{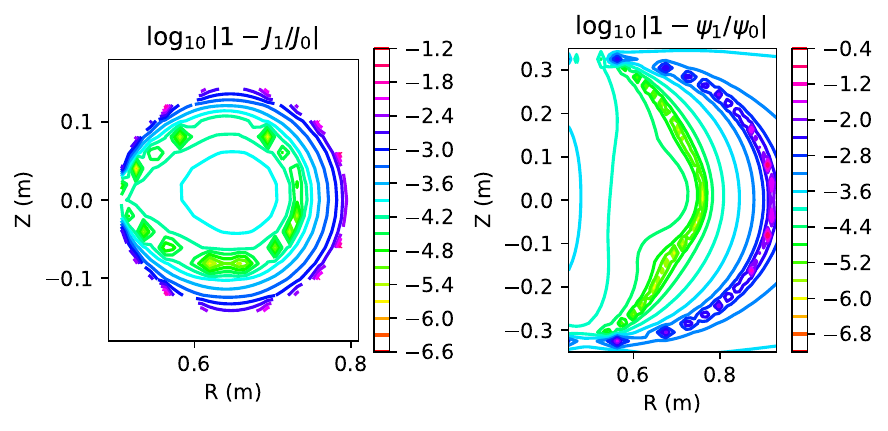}
\end{center}
\caption{$\log_{10}$ of absolute relative error in $J_\phi$ and $\psi$ from IPREQ and pyIPREQ for ADITYA-U case.}
\label{fig:validADITYA1}
\end{figure}

\begin{figure}[H]
\begin{center}
\includegraphics[width=\columnwidth]{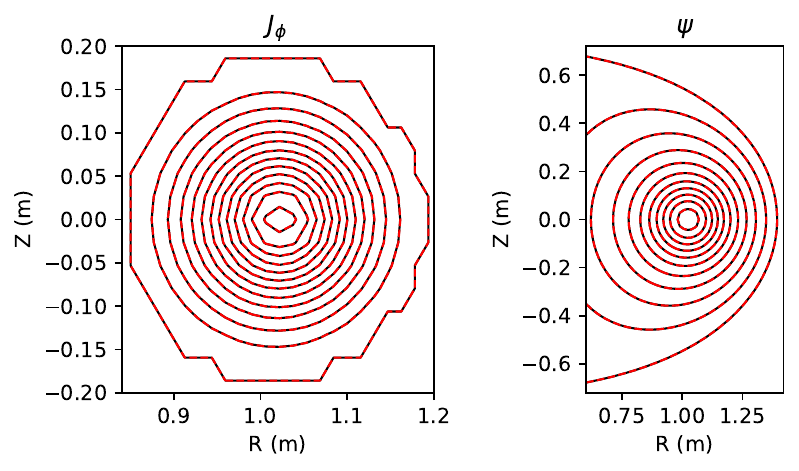}
\end{center}
\caption{Comparison of $J_\phi$ and $\psi$ from IPREQ ({\color{black}\boldlongemdash}) and pyIPREQ ({\color{red}\textbf{-\,-}}) for SST-1 case.}
\label{fig:validSST10}
\end{figure}

\begin{figure}[H]
\begin{center}
\includegraphics[width=\columnwidth]{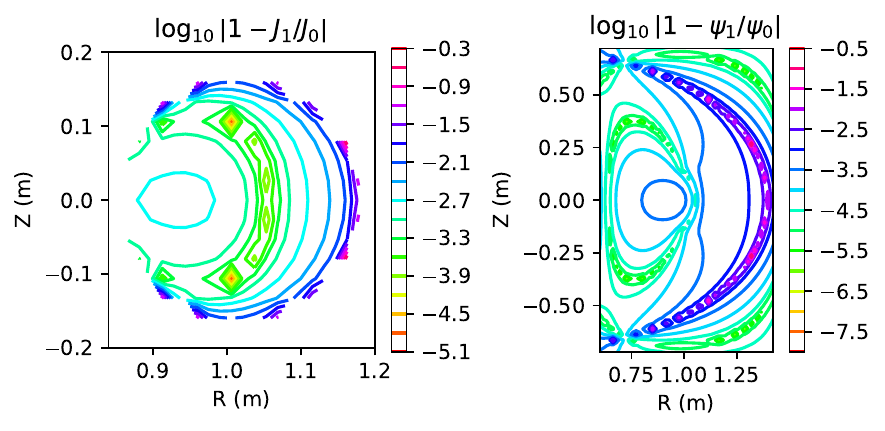}
\end{center}
\caption{$\log_{10}$ of absolute relative error in $J_\phi$ and $\psi$ from IPREQ and pyIPREQ for SST-1 case.}
\label{fig:validSST11}
\end{figure}

$J_\phi$ and $\psi$ obtained from IPREQ and pyIPREQ for ADITYA-U case are presented in figure \ref{fig:validADITYA0} and log of absolute relative differences are shown in figure \ref{fig:validADITYA1}. Magnetic axis values obtained from both codes for ADITYA-U case are shown in table \ref{tbl:ADITYAmagaxs}.
\begin{table}[H]
\begin{center}
\begin{tabular}{ l|l|l }
 \hline
 & IPREQ & pyIPREQ \\
 \hline
$R_{ax}$ & \SI{0.660289891}{\meter} & \SI{0.660289963}{\meter} \\
$Z_{ax}$ & \SI{0.0}{\meter} & \SI{-4.70e-05}{\meter} \\
$\psi_{ax}$ & \SI{0.01170516}{\tesla\meter^2} & \SI{0.01170589}{\tesla\meter^2} \\
 \hline
\end{tabular}
\end{center}
\caption{Magnetic axis position and $\psi$ obtained from IPREQ and pyIPREQ for ADITYA-U case.}
\label{tbl:ADITYAmagaxs}
\end{table}
$J_\phi$ and $\psi$ obtained from IPREQ and pyIPREQ for SST-1 case are presented in figure \ref{fig:validSST10} and log of absolute relative differences are shown in figure \ref{fig:validSST11}. Magnetic axis values obtained from both codes for SST-1 case are shown in table \ref{tbl:SSTmagaxs}.
\begin{table}[H]
\begin{center}
\begin{tabular}{ l|l|l }
 \hline
 & IPREQ & pyIPREQ \\
 \hline
$R_{ax}$ & \SI{1.0250}{\meter} & \SI{1.0251}{\meter} \\
$Z_{ax}$ & \SI{0.0}{\meter} & \SI{-4.97e-09}{\meter} \\
$\psi_{ax}$ & \SI{0.032698}{\tesla\meter^2} & \SI{0.032691}{\tesla\meter^2} \\
 \hline
\end{tabular}
\end{center}
\caption{Magnetic axis position and $\psi$ obtained from IPREQ and pyIPREQ for SST-1 case.}
\label{tbl:SSTmagaxs}
\end{table}
\subsubsection{Comparison with FREEGS}
\begin{figure}[H]
\begin{center}
\includegraphics[width=0.9\columnwidth]{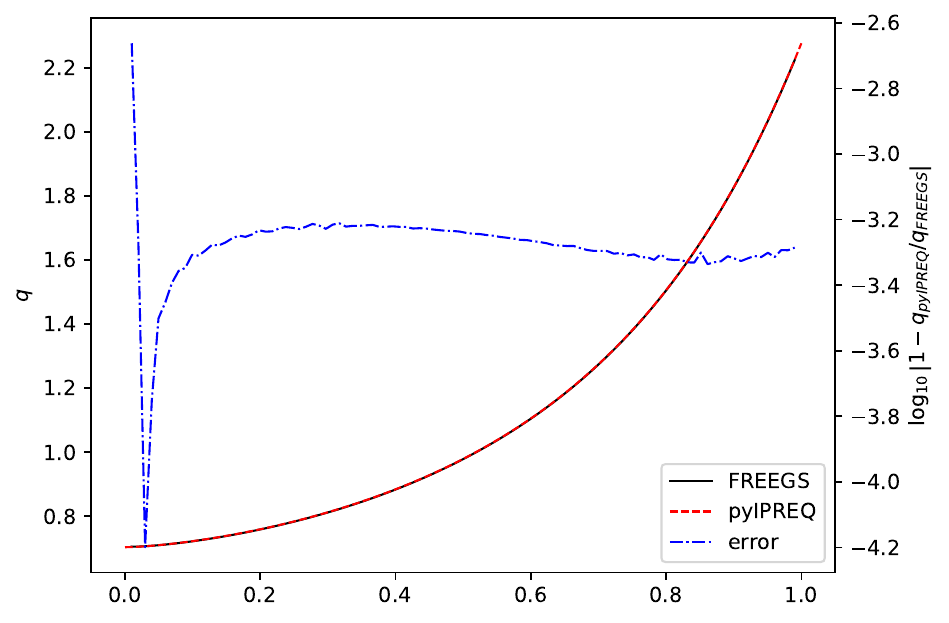}
\end{center}
\caption{Safety factor $q$ from FREEGS and pyIPREQ (left axis) and $\log_{10}$ of absolute relative error (right axis).}
\label{fig:FREEGSq}
\end{figure}

pyIPREQ code can write the output in G-EQDSK format \cite{1997Lao} and compute additional physical quantities like safety factor $q$, plasma beta $\beta$ and internal inductance $\ell_i$. Their definitions and benchmarking of implementation have been obtained from FREEGS code \cite{2024Ben}. For validating the implementation, an output G-EQDSK file was generated from pyIPREQ code for a hypothetical ADITYA-U case and the same file was loaded in the FREEGS code, then these quantities were re-computed using FREEGS. The comparison of plasma beta and internal inductance is shown in table \ref{tbl:FREEGSvars} and comparison of safety factor is shown in figure \ref{fig:FREEGSq}. The minor differences in all these results are likely due to some different algorithms used among them.
\begin{table}[H]
\begin{center}
\begin{tabular}{ l|l|l }
 \hline
 & FREEGS & pyIPREQ \\
 \hline
$\beta_p$ & $0.3539$ & $0.3527$ \\
$\beta_t$ & $0.005700$ & $0.005565$ \\
$\ell_i$ & $1.1108$ & $1.1147$ \\
 \hline
\end{tabular}
\end{center}
\caption{Plasma beta and internal inductance obtained from FREEGS and pyIPREQ.}
\label{tbl:FREEGSvars}
\end{table}
\section{Plasma Position Constrained MHD equilibrium}
\label{sec:RZaxcon}
To solve Grad-Shafranov equation \ref{eq:gs} constrained to given magnetic axis position $\left(R_{ax},Z_{ax}\right)$, an optimization based algorithm has been used in pyIPREQ, as presented in section \ref{sec:algo1}. In addition, an alternate geometric transformation based algorithm is also presented for $Z_{ax}$ in section \ref{sec:algo1Z}.
\subsection{Algorithm}
\label{sec:algo1}
Given the magnetic axis position $\left(R_{ax},Z_{ax}\right)$, the profile parameters $\alpha$, $\beta$ and $\gamma$, mentioned in equation \ref{eq:Jprofile}, are determined in two steps. First, $\beta$ is determined so that $J_\phi$ has maxima at $\left(R_{ax},Z_{ax}\right)$, i.e.,
\begin{equation}
\frac{\partial J_\phi}{\partial R}\bigg|_{\left(R_{ax},Z_{ax}\right)}=0
\label{eq:dJdRax}
\end{equation}
Sometimes during the iterations, $\beta$ may fall outside the acceptable range, in which case it can be clamped within a suitable subrange (like the one mentioned in \cite{2004Portone}), provided that the final $\beta$ remains physically reasonable. In the next step, $\alpha$ and $\gamma$ are determined using an optimization algorithm, similar to the other reconstruction codes mentioned in section \ref{sec:intro}. The objective is to minimize:
\begin{equation}
\chi=\sqrt{\left(\frac{R_{\nabla\psi=0}-R_{ax}}{\delta R_{ax}}\right)^2+\left(\frac{Z_{\nabla\psi=0}-Z_{ax}}{\delta Z_{ax}}\right)^2}
\label{eq:chi}
\end{equation}
where $\left(R_{\nabla\psi=0},Z_{\nabla\psi=0}\right)$ is the location of extremum of $\psi$ within limiter configuration and $\delta R_{ax},\delta Z_{ax}$ represent the measurement errors. It is important to note that the equation \ref{eq:dJdRax} only ensures that the radial location of the maximum of $J_\phi$ aligns with $R_{ax}$, but does not guarantee its alignment with the magnetic axis position (the location of maximum of $\psi$). However, enforcing this condition has been found to facilitate convergence of the optimization process. As with most optimization algorithms, the choice of initial parameter values significantly affects convergence and outcome. A practical strategy to determine a good initial guess is to pre-compute a map of $\chi$ over a broad range of $\alpha$ and $\gamma$ values before executing the full code.
\subsection{Test Case for Sensitivity}
\label{sec:testRax}
To test the sensitivity of the algorithm's convergence mentioned in section \ref{sec:algo1}, a test case was developed based on the same SST-1 parameters mentioned in section \ref{sec:IPREQcomp} but using the $J_\phi$ profiles from equation \ref{eq:Jprofile}. First, $J_\phi$ and $\psi$ were computed with profile parameters $\alpha=1.2$, $\beta=0.6$ and $\gamma=1.5$ as shown in figure \ref{fig:testJ0psi}:
\begin{figure}[H]
\begin{center}
\includegraphics[width=\columnwidth]{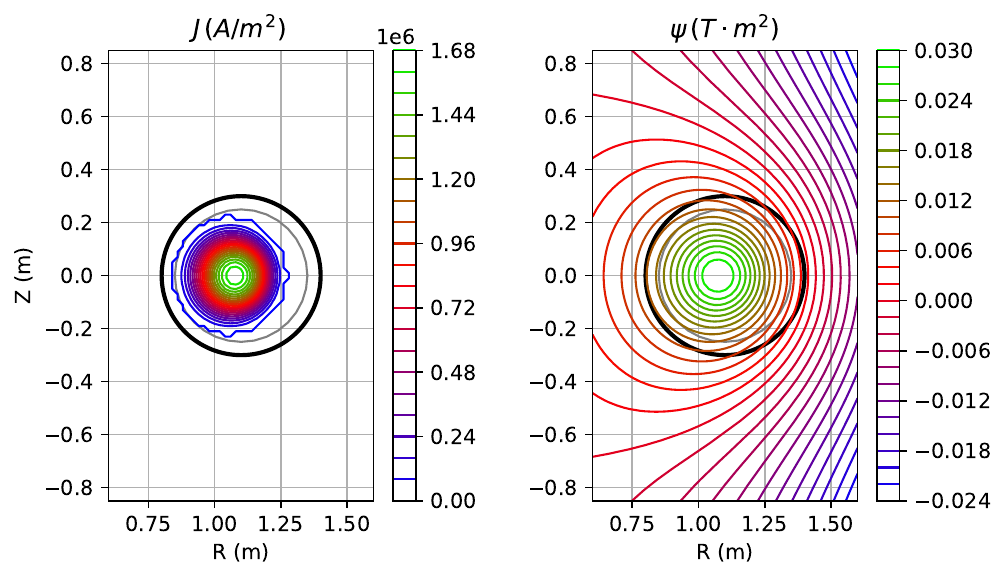}
\end{center}
\caption{$J_\phi$ and $\psi$ for SST-1 test case.}
\label{fig:testJ0psi}
\end{figure}
A direct test of section \ref{sec:algo1} algorithm is highly unlikely to yield same physical parameters because $\beta$ is systematically calculated by fixing the position of $J_\phi$ to the given $R_{ax}$ and then $\alpha,\gamma$ are estimated through optimization. Furthermore, a test case may also have different positions of maxima for $J_\phi$ and $\psi$. So instead, a partial test of the algorithm was performed by keeping the $\beta$ fixed instead of calculating it using equation \ref{eq:dJdRax}. In this manner, the optimization step was executed with different but close starting $\alpha_0,\gamma_0$ and the resultant $\alpha_1,\gamma_1$  values are shown in table \ref{tbl:paramstest}. These results show that only in 4 out of 9 test cases that the final parameters were close to the original values. This suggests that optimization is sensitive to initial guess values of parameters, even though a good convergence is achieved.
\begin{table}[H]
\begin{center}
\begin{tabular}{ l|l|l|l|l }
 \hline
$\alpha_0$ & $\gamma_0$ & $\alpha_1$ & $\gamma_1$ & $\chi$ \\
\hline
\num{1.2} & \num{1.5} & \num{1.207} & \num{1.505} & \num{6.3e-09} \\
\num{1.3} & \num{1.5} & \num{1.282} & \num{1.567} & \num{6.5e-09} \\
\num{1.3} & \num{1.6} & \num{1.333} & \num{1.609} & \num{4.8e-08} \\
\num{1.2} & \num{1.6} & \num{1.270} & \num{1.556} & \num{1.5e-09} \\
\num{1.1} & \num{1.6} & \num{1.180} & \num{1.484} & \num{4.8e-09} \\
\num{1.1} & \num{1.5} & \num{1.150} & \num{1.460} & \num{2.2e-08} \\
\num{1.1} & \num{1.4} & \num{1.093} & \num{1.417} & \num{1.4e-08} \\
\num{1.2} & \num{1.4} & \num{1.207} & \num{1.505} & \num{4.7e-08} \\
\num{1.3} & \num{1.4} & \num{1.199} & \num{1.499} & \num{3.3e-08} \\
\hline
\end{tabular}
\end{center}
\caption{Starting profile parameters $\alpha_0, \gamma_0$ and final profile parameters $\alpha_1, \gamma_1$ (rounded to 3 digits after decimal), for sensitivity test.}
\label{tbl:paramstest}
\end{table}
\subsection{Results for ADITYA-U Experiments}
\label{sec:results1}
\begin{figure}[H]
\begin{center}
\includegraphics[width=\columnwidth]{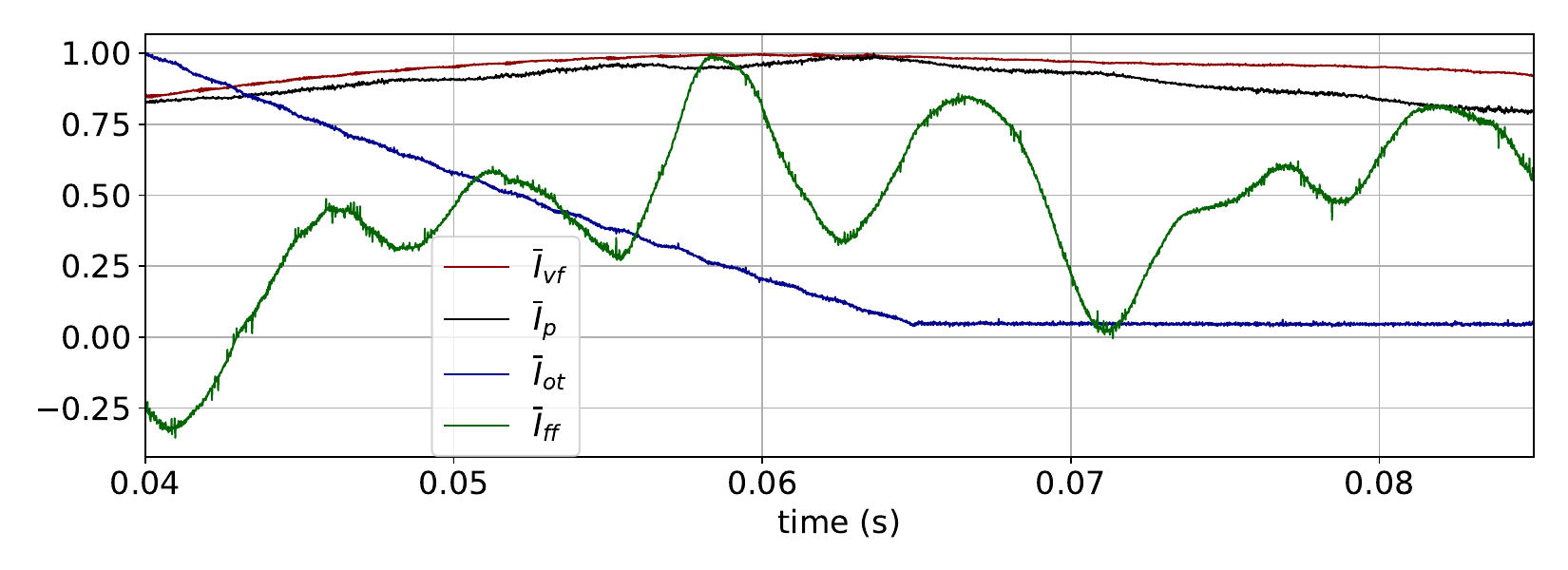}
\end{center}
\caption{Plasma current  and Coil currents for experiment 37681; $I_{p}=\bar I_{p}\SI{150.91}{kA}$, $I_{vf}=\bar I_{vf}\SI{5.4653}{kA}$/turn, $I_{vf}=\bar I_{vf}\SI{2.5276}{kA}$/turn, $I_{ff}=\bar I_{ff}\SI{1.4795}{kA}$/turn}
\label{fig:37681currs}
\end{figure}

\begin{figure}[H]
\begin{center}
\includegraphics[width=\columnwidth]{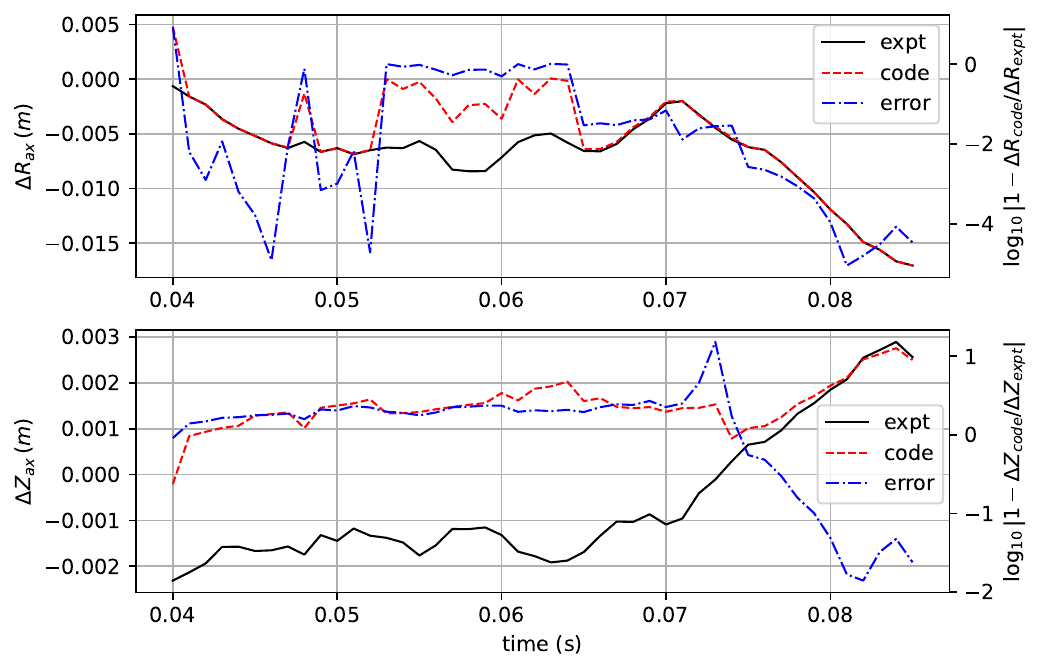}
\end{center}
\caption{$\Delta R_{ax}$ and $\Delta Z_{ax}$ from diagnostics and pyIPREQ for ADITYA-U experiment 37681, left axis shows actual values and right axis shows $\log_{10}$ of absolute relative error.}
\label{fig:37681RZax}
\end{figure}

\begin{figure}[H]
\begin{center}
\includegraphics[width=\columnwidth]{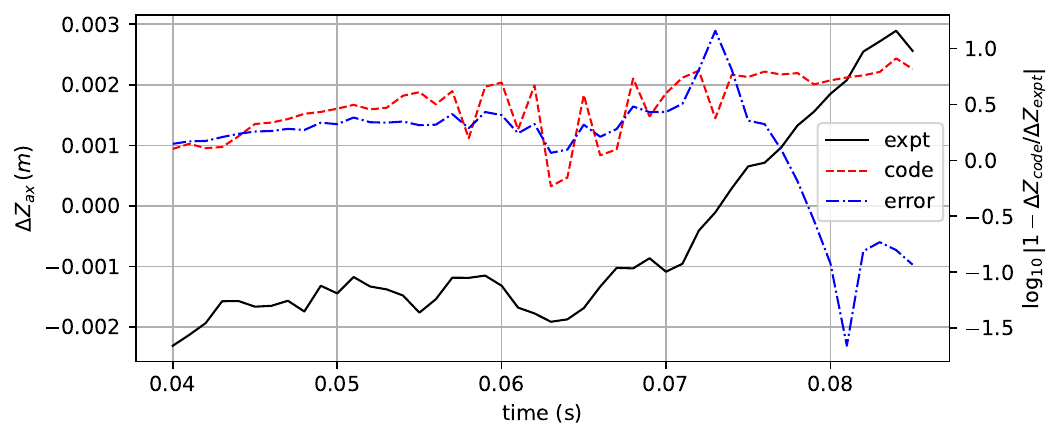}
\end{center}
\caption{$\Delta Z_{ax}$ from diagnostics and pyIPREQ (when only using $R$-term in $\chi$) for ADITYA-U experiment 37681, left axis shows actual values and right axis shows $\log_{10}$ of absolute relative error.}
\label{fig:37681Zax}
\end{figure}

\begin{figure}[H]
\begin{center}
\includegraphics[width=\columnwidth]{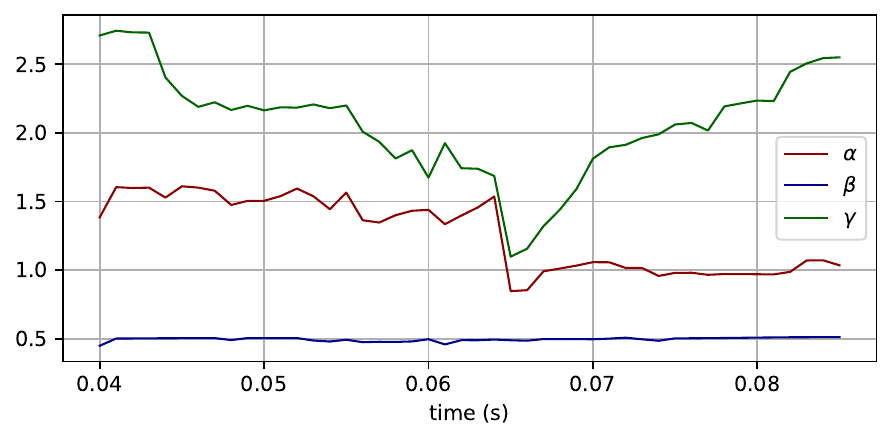}
\end{center}
\caption{$J_\phi$ profile parameters for ADITYA-U experiment 37681.}
\label{fig:37681abg}
\end{figure}

\begin{figure}[H]
\begin{center}
\includegraphics[width=\columnwidth]{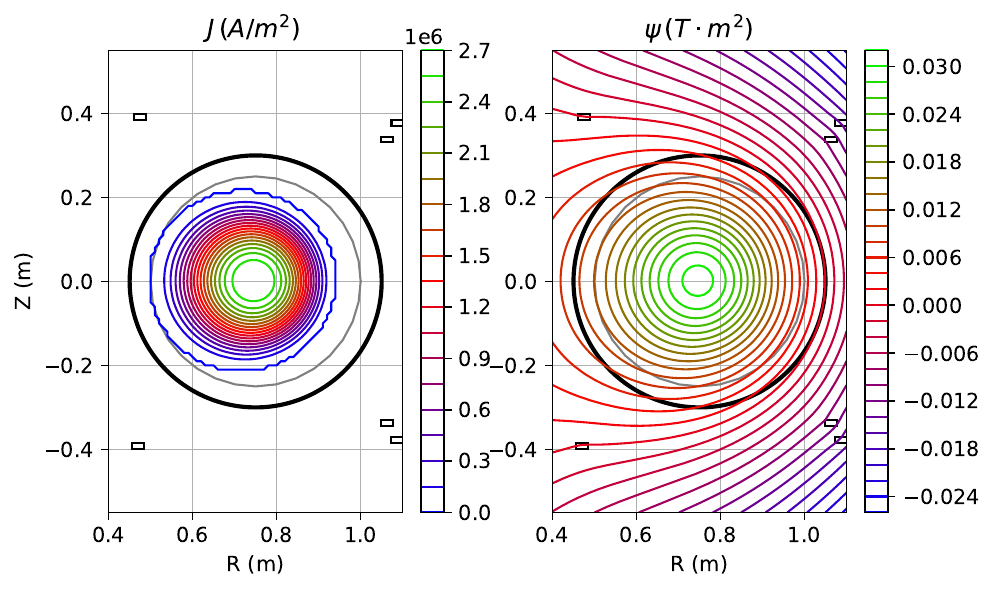}
\end{center}
\caption{$J_\phi$ and $\psi$ for ADITYA-U experiment 37681 at time \SI{0.06}{\second}.}
\label{fig:37681J0psi}
\end{figure}

\begin{figure}[H]
\begin{center}
\includegraphics[width=\columnwidth]{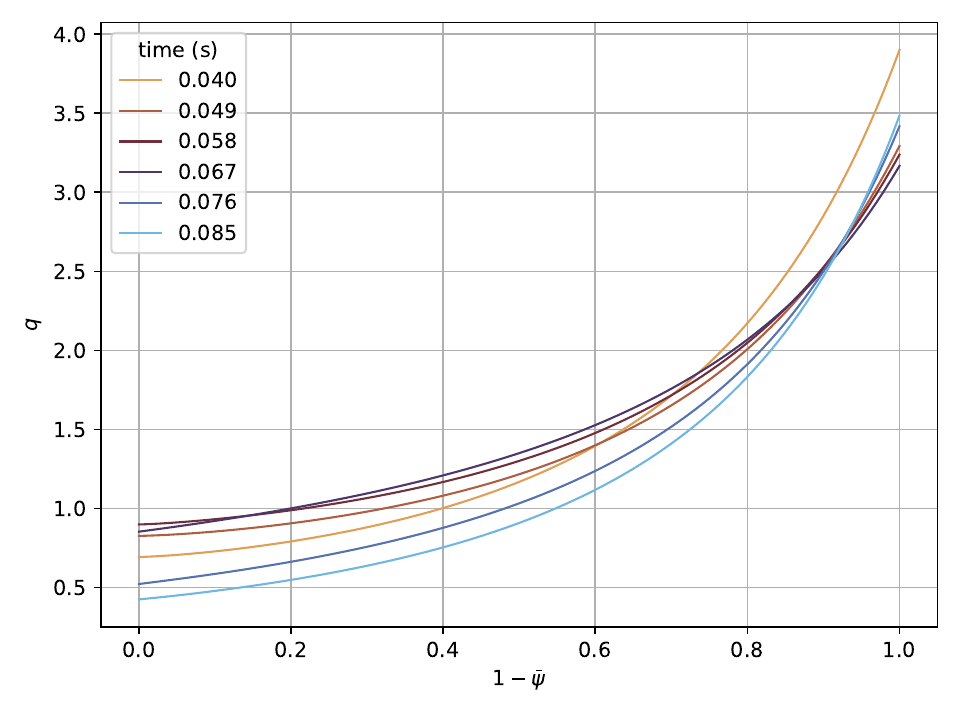}
\end{center}
\caption{Time evolution of plasma beta, internal inductance and safety factor for ADITYA-U experiment 37681.}
\label{fig:37681vars}
\end{figure}

\begin{figure}[H]
\begin{center}
\includegraphics[width=\columnwidth]{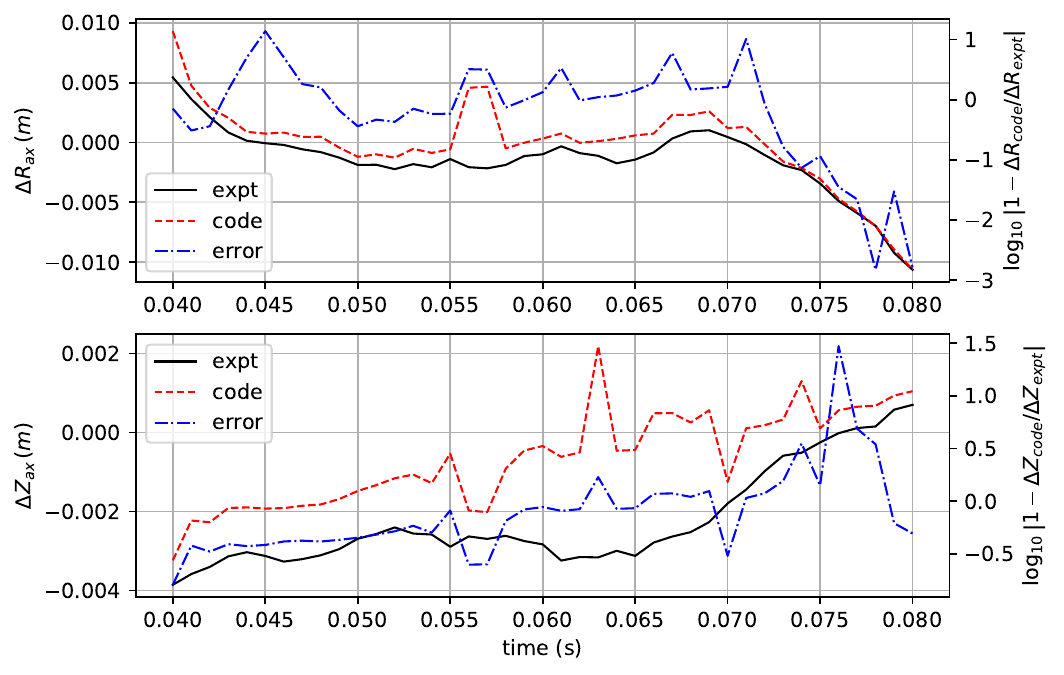}
\end{center}
\caption{$\Delta R_{ax}$ and $\Delta Z_{ax}$ for experiments 37012 and 37682 respectively. Values on left axis and error on right axis.}
\label{fig:moreRZax}
\end{figure}

pyIPREQ code has been applied to some plasma discharge experiments of ADITYA-U Tokamak, leveraging the availability of plasma current centroid position estimates from Sine-Cosine diagnostics \cite{2021Aich,2025Aich}, which is close to the magnetic axis position for circular low-$\beta$ plasma like ADITYA-U. The error estimates in both horizontal and vertical measurements are known to be about 7\% and 4\%, respectively \cite{2021Aich}. For each run, the code computes the MHD equilibrium over a time window that primarily covers the \emph{flat-top} phase of the discharge, during which the plasma remains relatively stable. The $R$-$Z$ grid spacing is set to \SI{1}{\cm}, and the time step is \SI{1}{\ms}. The ADITYA-U coil configuration used in the experiments is shown in table \ref{tbl:ADITYAcoils1}. For these results, some ADITYA-U poloidal field coils were ignored that are known to either cause or correct the error fields. Such error fields cannot be known without a more comprehensive analysis of the ADITYA-U Tokamak machine, like understanding the sources of eddy currents. Ignoring these coils might not reproduce an accurate MHD equilibrium for these experiments but is still useful to demonstrate the usefulness of the algorithms.
\begin{table}[H]
\footnotesize
\begin{center}
\begin{tabular}{ l|r|r|r|r|r }
 \hline
Coil & $R\,(m)$ & $Z\,(m)$ & $\Delta R\,(m)$ & $\Delta Z\,(m)$ & turns \\
 \hline
TR-1 & \num{0.22625} & \num{0} & \num{0.1075} & \num{1.04} & \num{174} \\
TR-2T & \num{0.3953} & \num{0.843} & \num{0.2305} & \num{0.16} & \num{56} \\
TR-2B & \num{0.3953} & \num{-0.83952} & \num{0.2305} & \num{0.157} & \num{56} \\
TR-3T & \num{1.22319} & \num{0.72806} & \num{0.03483} & \num{0.06} & \num{-3} \\
TR-3B & \num{1.226} & \num{-0.727} & \num{0.03516} & \num{0.0595} & \num{-3} \\
TR-4T & \num{1.534} & \num{0.60209} & \num{0.06446} & \num{0.0382} & \num{4} \\
TR-4B & \num{1.53007} & \num{-0.601} & \num{0.0654} & \num{0.03867} & \num{4} \\
TR-5T & \num{1.627} & \num{1.148} & \num{0.034} & \num{0.0155} & \num{-1} \\
TR-5B & \num{1.627} & \num{-1.1475} & \num{0.035} & \num{0.016} & \num{-1} \\
BV-1T & \num{0.37981} & \num{1.0527} & \num{0.16405} & \num{0.12003} & \num{-60} \\
BV-1B & \num{0.382} & \num{-1.051} & \num{0.16937} & \num{0.11733} & \num{-60} \\
BV-2T & \num{1.64204} & \num{1.1885} & \num{0.18155} & \num{0.03863} & \num{-22} \\
BV-2B & \num{1.641} & \num{-1.189} & \num{0.1795} & \num{0.038} & \num{-22} \\
MDO-T & \num{1.06288} & \num{0.3375} & \num{0.028} & \num{0.01312} & \num{1} \\
MDO-B & \num{1.06288} & \num{-0.3375} & \num{0.02814} & \num{0.01326} & \num{1} \\
FFI-T & \num{0.475} & \num{0.392} & \num{0.02943} & \num{0.0152} & \num{-1} \\
FFI-B & \num{0.47} & \num{-0.392} & \num{0.02923} & \num{0.01496} & \num{-1} \\
FFO-T & \num{1.0875} & \num{0.377} & \num{0.02814} & \num{0.01291} & \num{1} \\
FFO-B & \num{1.0875} & \num{-0.377} & \num{0.02834} & \num{0.01335} & \num{1} \\
 \hline
\end{tabular}
\end{center}
\caption{ADITYA-U poloidal field coils configuration used for results in section \ref{sec:results1}.}
\label{tbl:ADITYAcoils1}
\end{table}
In experiment 37681, the plasma current and coil currents in the flat-top region are shown in figure \ref{fig:37681currs}. $I_{ot}$ is applicable to all TR coils, $I_{vf}$ is applicable to all BV coils and $I_{ff}$ is applicable to the remaining coils mentioned in the table \ref{tbl:ADITYAcoils1}.
A comparison of the magnetic axis displacement $\Delta R_{ax}\equiv R_{ax}-R_0$ and $\Delta Z_{ax}\equiv Z_{ax}$ obtained from diagnostic data and pyIPREQ calculations is shown in figure \ref{fig:37681RZax}.
As shown in figure \ref{fig:37681RZax}, $\Delta R_{ax}$ shows a good agreement between experiment and simulation, but $\Delta Z_{ax}$ does not appear to match as good as $\Delta R_{ax}$. From observations of running the pyIPREQ code for several other ADITYA-U experiments, it appears that $\Delta Z_{ax}$ is more influenced by poloidal field coil currents and configuration than the $J_\phi$ profile parameters. Nonetheless, this algorithm still produces slightly improved results for $\Delta Z_{ax}$, as can be seen in figure \ref{fig:37681Zax}, which is obtained by using $\chi$ that only optimizes for $\Delta R_{ax}$.
Numerical convergence alone is insufficient to establish the usefulness of the code. It is also necessary to assess the physical viability of the resulting plasma profiles and their temporal evolution. For instance, an overly peaked $J_\phi$ profile centered at $\left(R_{ax},Z_{ax}\right)$ would resemble a single-filament model and may lack physical meaning. Another potential issue is the sensitivity of the profile parameters to their initial guesses, as demonstrated in section \ref{sec:testRax}, which can lead to abrupt and non-physical changes in the plasma behavior. For the results shown in figure \ref{fig:37681RZax}, the evolution of $J_\phi$ profile parameters is shown in figure \ref{fig:37681abg}. $\beta$ values around $0.5$ indicates a balance between pressure and poloidal current terms in equation \ref{eq:Jphi}. The reasonable values and mostly smooth variation of profile parameters indicate a physically viable evolution of plasma. Profiles of $J_\phi$ and $\psi$ at time \SI{0.06}{\second} are shown in figure \ref{fig:37681J0psi} and time evolution of physical parameters like plasma beta, internal inductance and safety factor is shown in figure \ref{fig:37681vars}.
\begin{center}
\includegraphics[width=\columnwidth]{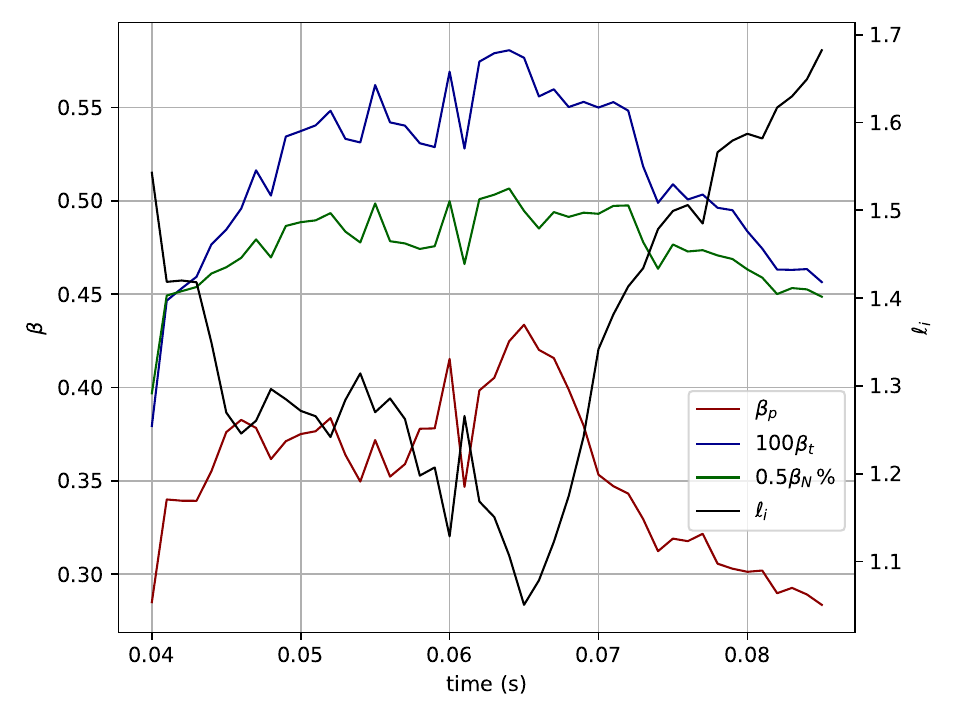}
\end{center}
Time evolutions of $\Delta R_{ax}$ and $\Delta Z_{ax}$ of some other ADITYA-U experiments, for which reasonably good output were obtained, are shown in figure \ref{fig:moreRZax}.
\begin{center}
\includegraphics[width=\columnwidth]{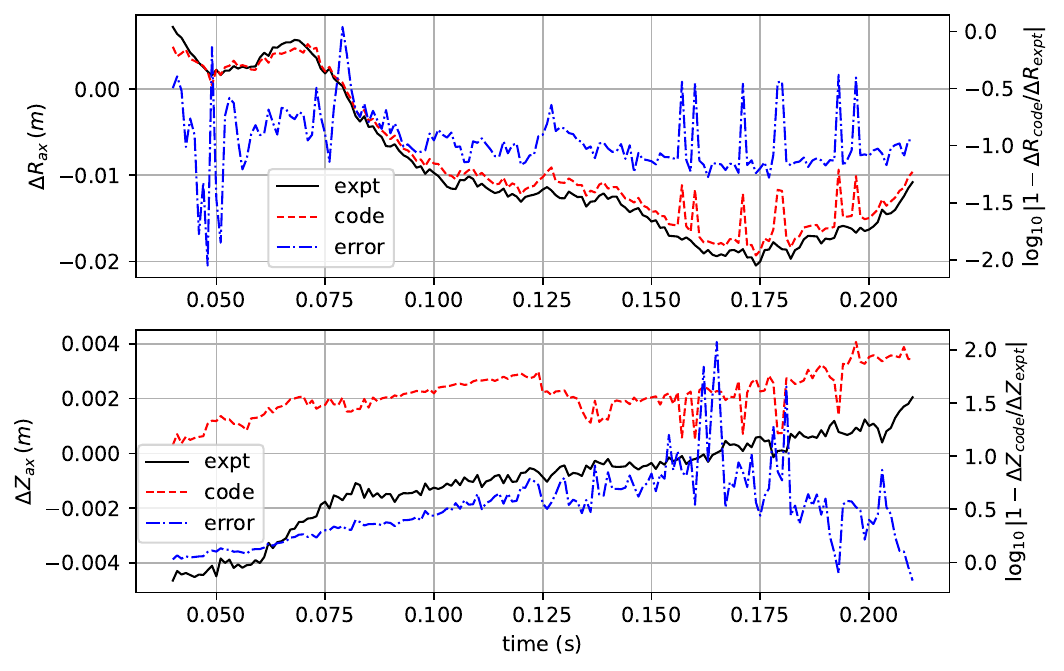}
\end{center}
\subsection{Alternate Algorithm for $\Delta Z_{ax}$}
\label{sec:algo1Z}
\begin{figure}[H]
\begin{center}
\includegraphics[width=0.75\columnwidth]{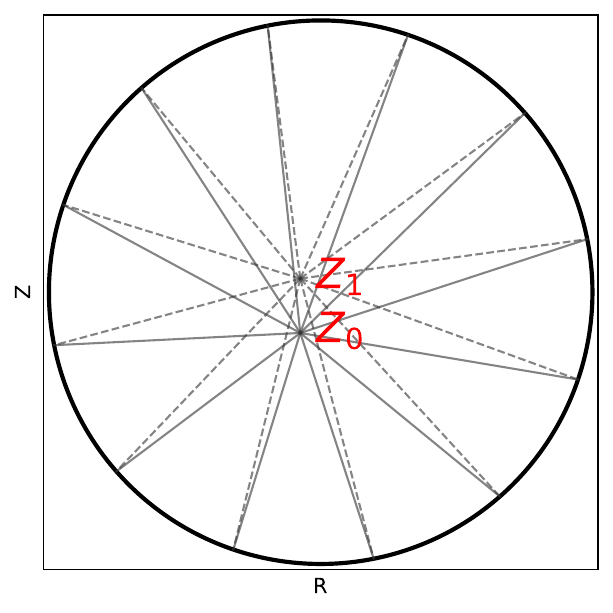}
\end{center}
\caption{Linear transformation applied to $J_\phi\left(R,Z\right)$ so that maxima $Z_0$ moves to $Z_1$.}
\label{fig:Zshift}
\end{figure}

\begin{figure}[H]
\begin{center}
\includegraphics[width=\columnwidth]{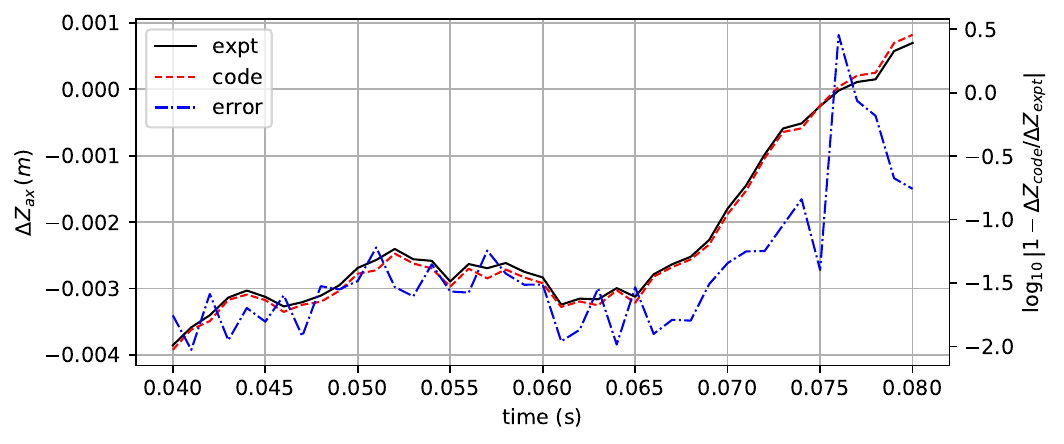}
\end{center}
\caption{$\Delta Z_{ax}$ from diagnostics and pyIPREQ (using $Z_{ax}$ shift algorithm) for ADITYA-U experiments 37681, 37012 and 37682 respectively. Values on left axes and error on right axes.}
\label{fig:ADITYAZax}
\end{figure}

\begin{figure}[H]
\vspace{-1.15cm}
\begin{center}
\includegraphics[width=\columnwidth]{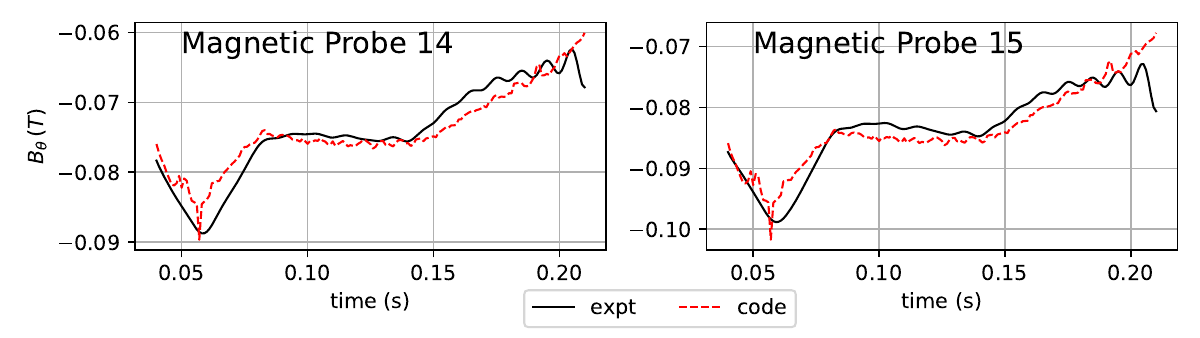}
\end{center}
\caption{Comparison of Magnetic Probes Data with results of Section \ref{sec:algo1Z}, for ADITYA-U experiment 37012.}
\label{fig:37012mc}
\end{figure}

From the discussion and the results of section \ref{sec:results1}, it appears that the optimization algorithm is not good enough to constrain $Z_{ax}$. In this section, an alternate geometric transformation based algorithm is presented that can produce better alignment of $Z_{ax}$ with desired values. In this algorithm, during iterations the boundary of $J_\phi$ is kept fixed, while a linear transformation is internally applied to $J_\phi$ to shift the vertical position of its maximum from $Z_0$ to $Z_1$ (where $Z_1$ is the prescribed $Z_{ax}$), as illustrated in figure \ref{fig:Zshift}. In this process, values of $J_\phi$ located along the solid black lines are moved to the corresponding dashed black lines. $J_\phi$ is then interpolated onto the $R$-$Z$ grid, and the total magnitude is adjusted with plasma current $I_p$. This approach has generally been observed to be compatible with the Picard iterations and the optimization algorithm.
A similar approach, involving vertically shifting the entire $J_\phi$ distribution (including its boundary), was also attempted but was found to cause failures during the iteration process. It should be noted that such geometric transformations are not fully consistent with the form of equation \ref{eq:Jphi}, even if Picard iterations converge. Nevertheless, when the applied vertical shift reaches zero or becomes negligible in the final iteration, the resulting equilibrium solutions can still be acceptable. Results for this method, combined with a constrained $R_{ax}$, are shown in figure \ref{fig:ADITYAZax}. In each case, the vertical shift at the final iteration was found to be less than \SI{1}{\mm} at each timestep (except for initial time durations of experiment 37012). $\Delta R_{ax}$ obtained in these results were mostly same as shown in the section \ref{sec:results1} and thus are not shown here.
\begin{center}
\includegraphics[width=\columnwidth]{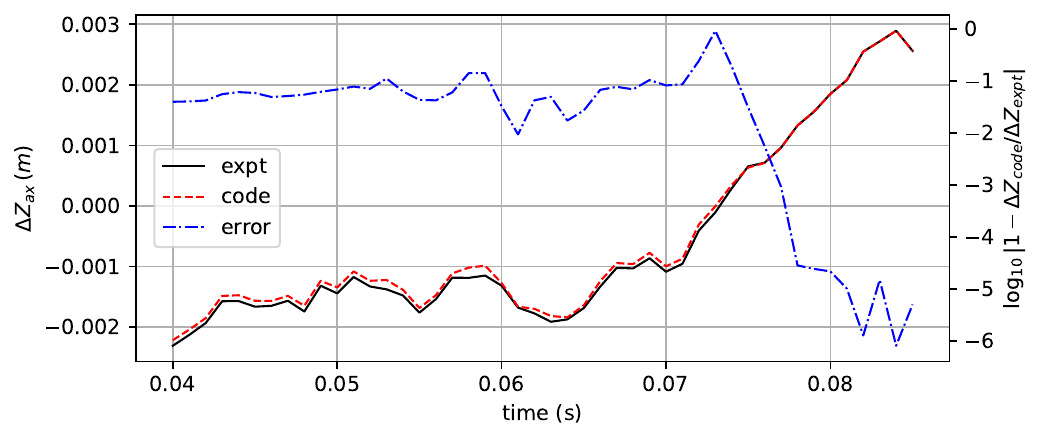}
\end{center}
\begin{center}
\includegraphics[width=\columnwidth]{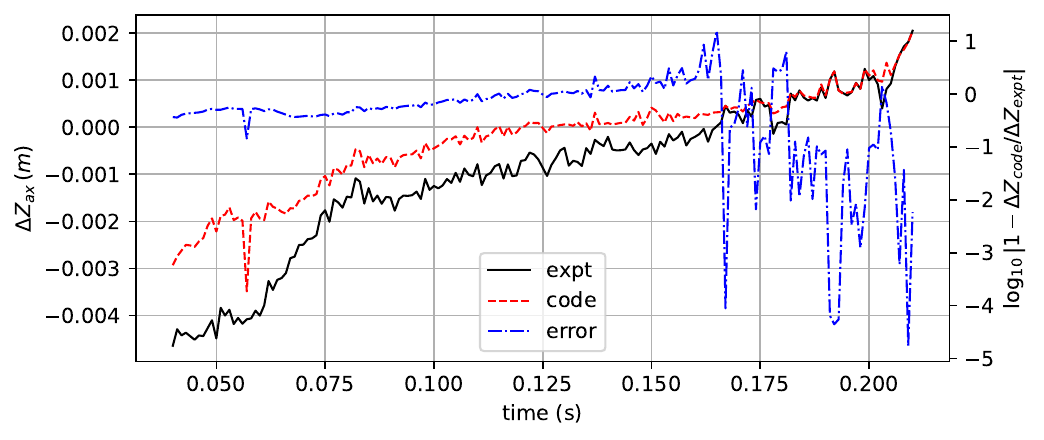}
\end{center}
The results presented in this section and in Section \ref{sec:results1} do not constitute a full MHD equilibrium reconstruction, which requires including more diagnostics data and sophisticated $J_\phi$ profile. However, these results can still provide a sufficiently accurate description in certain simple cases, such as the circular plasma of ADITYA-U. To illustrate this, figure \ref{fig:37012mc} shows a comparison for ADITYA-U Experiment 37012, where magnetic probe data are available. In ADITYA-U Tokamak, 16 probes are arranged uniformly on a circular contour of radius \SI{27.5}{\cm} in a poloidal plane, providing $B_\theta$ measurements \cite{2021Aich}. Reliable data were obtained from 14 of these probes, as the signals from probes 9 and 16 were unavailable. Using the $\psi$ output from the reconstruction results described in this section, the computed $B_\theta$ values exhibit good agreement with the probe measurements, as shown in figure \ref{fig:37012mc}. This demonstrates that even a position-constrained reconstruction can reproduce the essential features of a simple circular plasma.
\begin{center}
\includegraphics[width=\columnwidth]{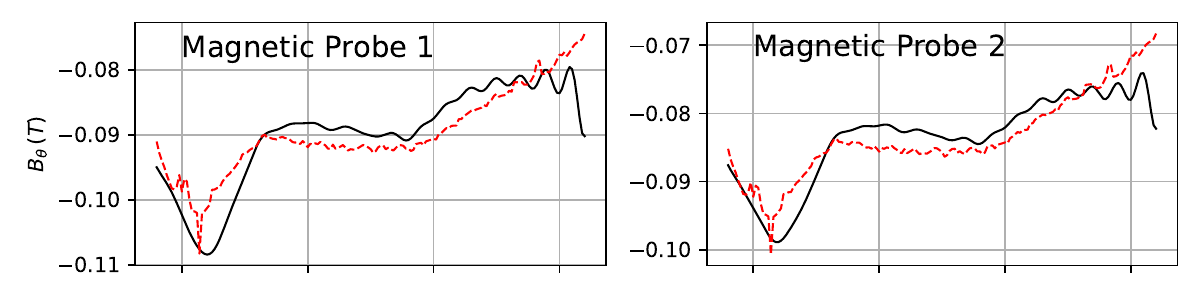}
\includegraphics[width=\columnwidth]{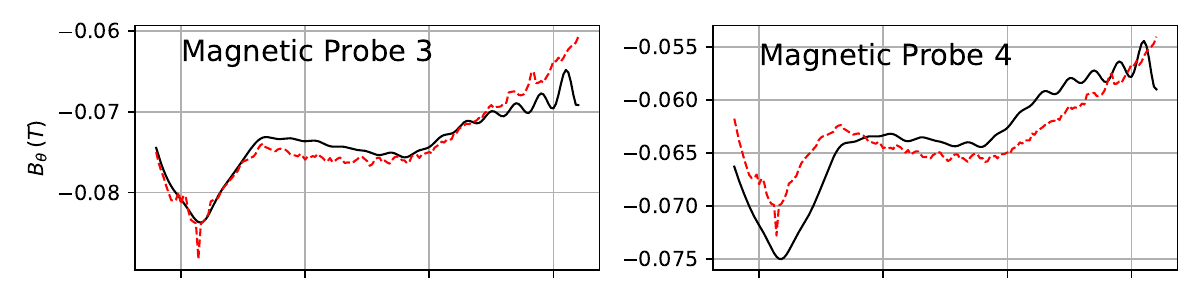}
\includegraphics[width=\columnwidth]{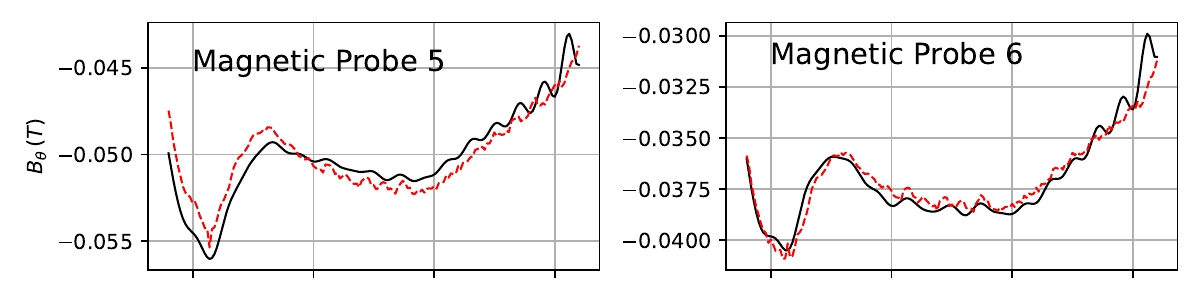}
\includegraphics[width=\columnwidth]{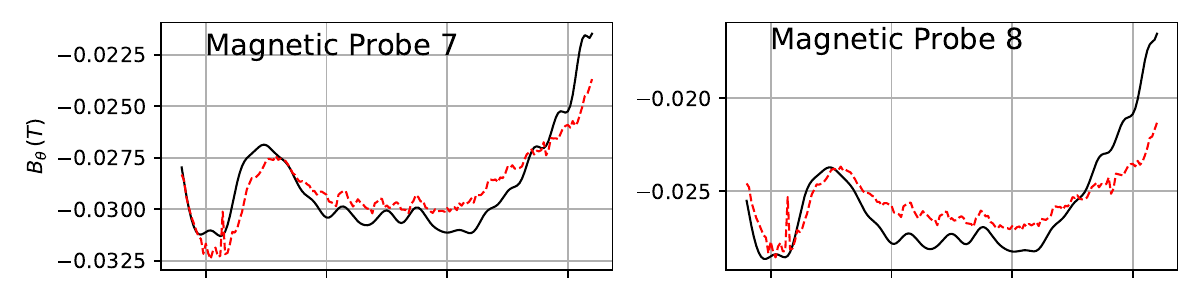}
\includegraphics[width=\columnwidth]{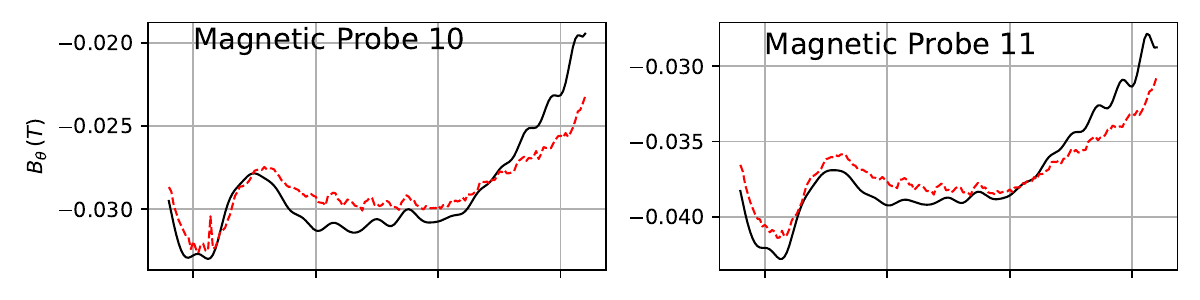}
\includegraphics[width=\columnwidth]{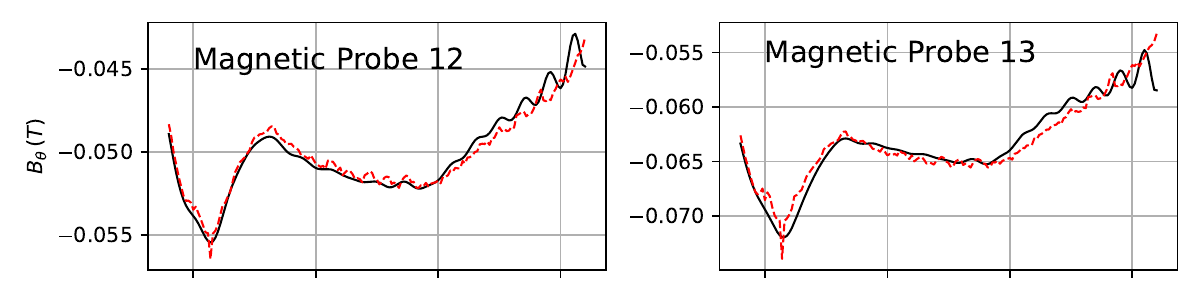}
\end{center}
\section{Predictions for SST-1 and ADITYA-U}
The algorithm and results mentioned in section \ref{sec:RZaxcon} can be used in several ways like generating MHD equilibrium scenarios or predicting differences in output with minor changes in inputs. In section \ref{sec:predsst1u}, two plasma scenarios are generated for proposed SST-1 future upgrades and in section \ref{sec:predadityau}, differences are presented in ADITYA-U experiment if divertor coils were used.
\subsection{SST-1 Plasma Scenarios}
\label{sec:predsst1u}
\begin{figure}[H]
\begin{center}
\includegraphics[width=0.5\columnwidth]{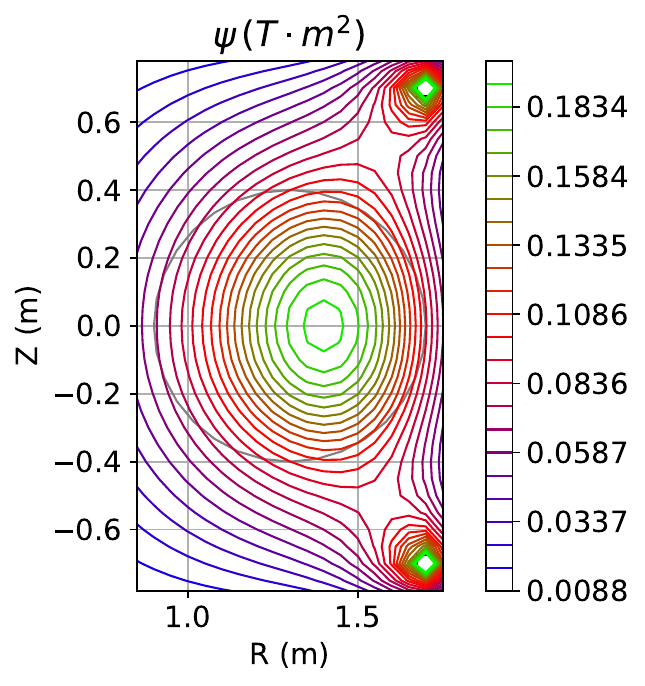}
\includegraphics[width=0.48\columnwidth]{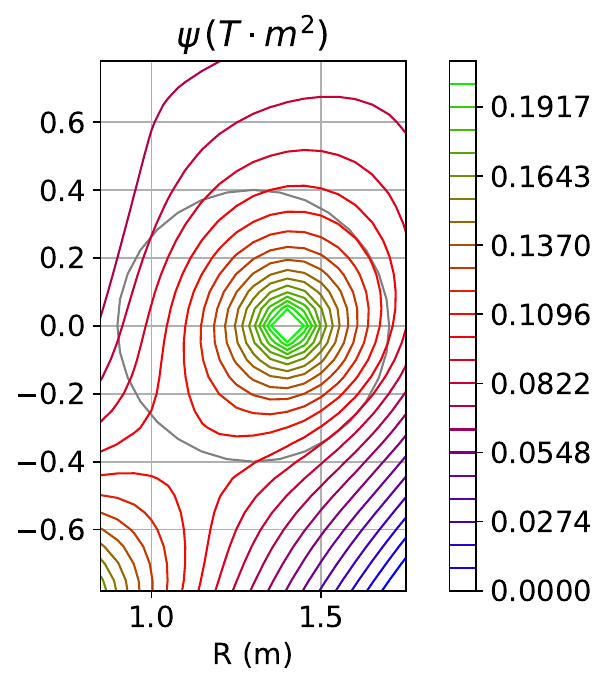}
\end{center}
\caption{Negative Triangularity (left) and Single-Null (right) Plasma Scenarios for SST-1 proposed upgrades.}
\label{fig:sst1u}
\end{figure}

In this section, potential plasma scenarios are generated for proposed SST-1 future upgrades. Instead of guess $J_\phi$ profile parameters, desired magnetic axis position is used with constrained plasma position algorithm. The flat-top scenarios are generated for a negative triangularity case (Case 1) and a single null case (Case 2), using newly proposed poloidal field coils. The machine parameters, poloidal field coil configuration and currents used to generate these are shown in tables \ref{tbl:SST1Uparams}, \ref{tbl:SST1Ucoils} and \ref{tbl:SST1Ucurrs}. The resultant flux surface plots are shown in figure \ref{fig:sst1u}.
\begin{table}[H]
\begin{center}
\begin{tabular}{ l|r }
 \hline
$R_0$, $a$ & \SI{1.3}{\meter}, \SI{0.4}{\meter} \\
$I_p$ & \SI{200}{\kA} (Case 1), \SI{450}{\kA} (Case 2) \\
$R_{ax}, Z_{ax}$ &\SI{1.4}{\meter}, \SI{0.0}{\meter} \\
 \hline
\end{tabular}
\end{center}
\caption{Machine and Profile Parameters used in case of SST-1 proposed upgrades.}
\label{tbl:SST1Uparams}
\end{table}
\begin{table}[H]
\footnotesize
\begin{center}
\begin{tabular}{ l|r|r|r|r|r }
 \hline
Coil & $R\,(m)$ & $Z\,(m)$ & $\Delta R\,(m)$ & $\Delta Z\,(m)$ & turns \\
 \hline
PF3 & \num{0.26} & $\pm$\num{0.85} & \num{0.154} & \num{0.11} & \num{35} \\
PF5 & \num{2.545} & $\pm$\num{1.485} & \num{0.162} & \num{0.162} & \num{36} \\
PF6 & \num{1.7} & $\pm$\num{0.7} & \num{0.05} & \num{0.05} & \num{10} \\
 \hline
\end{tabular}
\end{center}
\caption{Poloidal field coils configuration used in case of SST-1 proposed upgrades.}
\label{tbl:SST1Ucoils}
\end{table}
\begin{table}[H]
\begin{center}
\begin{tabular}{ l|r|r }
 \hline
&  \multicolumn{2}{c}{$I_{\text{coil}}$ (\SI{}{kA/turn})} \\
 \hline
Coil & Case 1 & Case 2 \\
 \hline
PF3-T & \num{0} & \num{28.8185} \\
PF3-B & \num{0} & \num{80.9258} \\
PF5-T & \num{-21.1139} & \num{-4.5360} \\
PF5-B & \num{-21.1139} & \num{-10.9859} \\
PF6-T & \num{26.4538} & \num{0} \\
PF6-B & \num{26.4538} & \num{0} \\
 \hline
\end{tabular}
\end{center}
\caption{Poloidal field coils currents used in case of SST-1 proposed upgrades.}
\label{tbl:SST1Ucurrs}
\end{table}
\subsection{ADITYA-U with Divertor Coils}
\label{sec:predadityau}
\begin{figure}[H]
\begin{center}
\includegraphics[width=\columnwidth]{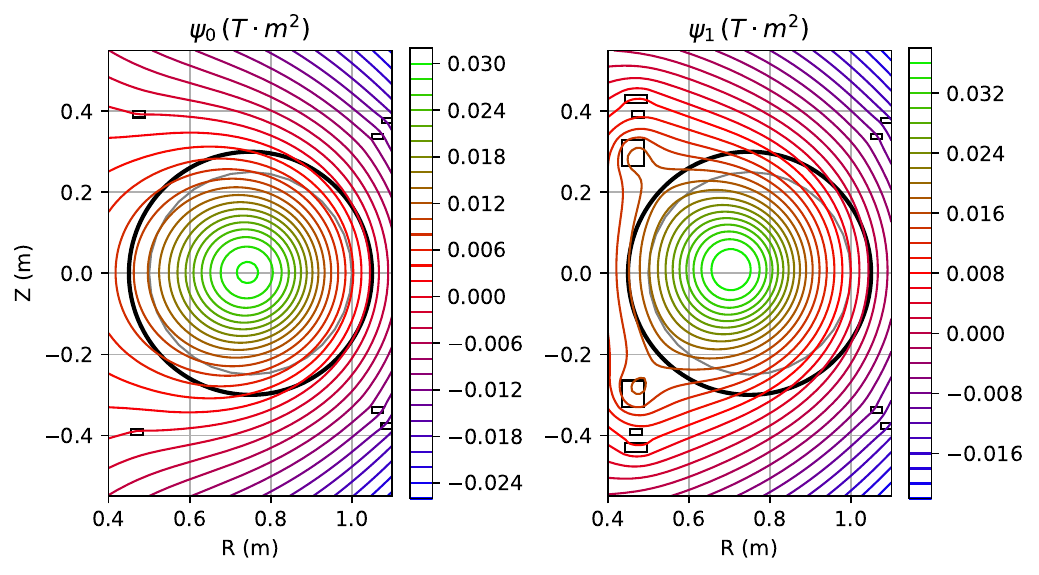}
\end{center}
\caption{$\psi$ at \SI{0.06}{\second} without divertor coils (left) and with divertor coils (right).}
\label{fig:37681psidiv}
\end{figure}

\begin{figure}[H]
\begin{center}
\includegraphics[width=\columnwidth]{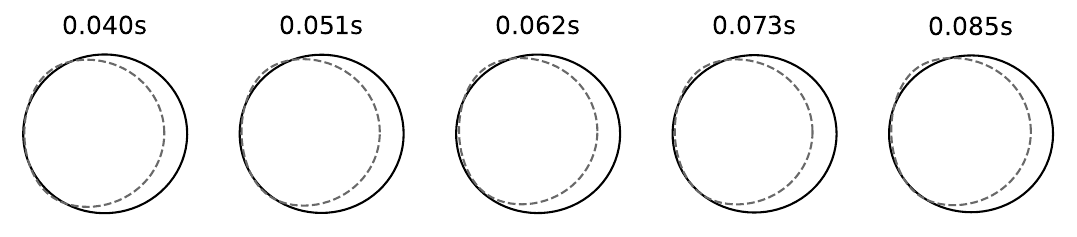}
\end{center}
\caption{Evolution of LCFS without divertor coils ({\color{black}\boldlongemdash}) and with divertor coils ({\color{gray}\textbf{-\,-}}).}
\label{fig:37681lcfsdiv}
\end{figure}

ADITYA-U is equipped with divertor coils that can be used to shape the plasma configuration. In this section, a comparison is presented based on the results from section \ref{sec:results1}, illustrating the predicted impact on the flux surfaces and Last Closed Flux Surfaces (LCFS) if these divertor coils were active during experiment 37681. The specification of these additional divertor coils is shown in table \ref{tbl:ADITYAcoils2}.
\begin{table}[H]
\footnotesize
\begin{center}
\begin{tabular}{ l|r|r|r|r|r }
 \hline
Coil & $R\,(m)$ & $Z\,(m)$ & $\Delta R\,(m)$ & $\Delta Z\,(m)$ & turns \\
 \hline
MDI-T & \num{0.4625} & \num{0.297} & \num{0.055} & \num{0.065} & \num{12} \\
MDI-B & \num{0.462} & \num{-0.297} & \num{0.055} & \num{0.065} & \num{12} \\
ADI-T & \num{0.47} & \num{0.43} & \num{0.055} & \num{0.02166} & \num{4} \\
ADI-B & \num{0.4704} & \num{-0.43} & \num{0.055} & \num{0.02166} & \num{4} \\
 \hline
\end{tabular}
\end{center}
\caption{ADITYA-U divertor coils configuration used in section \ref{sec:predadityau}.}
\label{tbl:ADITYAcoils2}
\end{table}
Consider a hypothetical scenario for this experiment, where the divertor coil current is ramped up from 0 to \SI{3}{\kA/turn} over the time interval from 0 to \SI{0.05}{\second}. The current is then held constant until \SI{0.09}{\second}, after which it is ramped down to 0 by \SI{0.14}{\second}. The $J_\phi$ profile parameters are assumed same as shown in figure \ref{fig:37681abg}. The resultant effect of these divertor coils on $\psi$ at \SI{0.06}{\second} is shown in figure \ref{fig:37681psidiv}, and the evolution of the LCFS during the discharge is shown in figure \ref{fig:37681lcfsdiv}. As seen from these figures, the plasma shape changes from circular to a more D-shaped configuration under the influence of these divertor coils.
\section{Conclusion}
\label{sec:conclusion}
A free-boundary MHD equilibrium code named pyIPREQ was developed based on the codes PEST and IPREQ. It was developed in Python-3 programming language and offers some new enhancements and features. The code has been successfully benchmarked with a published ITER case and the original IPREQ code for test cases of the ADITYA-U and SST-1 Tokamaks. The pyIPREQ code was then extended to compute MHD equilibrium constrained to prescribed magnetic axis position. This feature was tested for experimental data of ADITYA-U Tokamak where magnetic axis positions were inferred from the available diagnostics data. The results demonstrate that the code performs as expected in reconstructing equilibrium profiles. This code was then used to predict plasma scenarios for proposed upgrade of SST-1 and evolution of flux surfaces if divertor coils were used in a particular ADITYA-U experiment. \\

Among the shortcomings of the code, one is that its performance is critical to the initial guess values of profile parameters $\alpha$ and $\gamma$, which need to be carefully selected. To obtain meaningful results in experimental applications, the profile parameters should be within physical expectations and have a smooth temporal variation, this may require some trial and error approach with initial guess values. Additionally, the position-constrained algorithm enforces alignment between the maxima of $J_\phi$ and $\psi$, which is not strictly valid, as seen from equation \ref{eq:Jphi}. However, this approach has been found to be more effective than direct optimization of $\beta$. Another limitation of pyIPREQ is its computational efficiency. Since it is implemented in Python-3 without any parallelization or low-level optimizations, it is currently slower than the original IPREQ code, which was developed in FORTRAN. Future improvements could focus on optimizing the code for performance, incorporating additional diagnostic inputs such as flux loops and magnetic probes, and validating its applicability to other Tokamak experiments, particularly those with shaped plasma configurations.

\subsubsection*{Code and Data Availability}
The pyIPREQ code is currently available on request to the corresponding author. It will be made available and kept updated on the author's \href{https://github.com/udy11/}{GitHub: https://github.com/udy11/} and \href{https://gitlab.com/udy11/}{GitLab: https://gitlab.com/udy11/}.  The data that support the findings of this study are available from the corresponding author upon reasonable request.
\subsubsection*{Acknowledgment}
The author is thankful to Late Dr. R. Srinivasan for providing the original IPREQ code to develop the pyIPREQ code. The author is thankful to Dr. Olivier Sauter (Swiss Plasma Center - EPFL, Switzerland) for helping in figuring out the COCOS number for pyIPREQ. This work was supported by the Institute for Plasma Research (IPR), India.

\end{multicols}

\bibliographystyle{unsrt}
\bibliography{main}

\end{document}